%% file: DGTD_ACM_JCP.tex
\documentclass[final,3p,onecolumn,number,sort&compress,10pt,letterpaper]{elsarticle}
\pdfoutput=1
\input{header_jcp.tex}

\journal{Journal of Computational Physics}

\begin{document}

\begin{frontmatter}

\title{A Discontinuous Galerkin Time-Domain Method with Dynamically Adaptive Cartesian Meshes for Computational Electromagnetics}

\author[uiuc]{Su~Yan}
\ead{suyan@illinois.edu}

\author[uiuc]{Chao-Ping~Lin}
\ead{clin108@illinois.edu}

\author[cfdrc]{Robert~R.~Arslanbekov}
\ead{robert.arslanbekov@cfdrc.com}

\author[cfdrc]{Vladimir~I.~Kolobov}
\ead{vladimir.kolobov@cfdrc.com}

\author[uiuc]{Jian-Ming~Jin\corref{cor1}}
\ead{j-jin1@illinois.edu}

\cortext[cor1]{Corresponding author}
\address[uiuc]{Center for Computational Electromagnetics, Department of Electrical and Computer Engineering, University of Illinois at Urbana-Champaign, Urbana, IL 61801-2991, USA}
\address[cfdrc]{CFD Research Corporation, 701 McMillian Way, Suite D, Huntsville, AL 35806, USA}





\begin{abstract}
  A discontinuous Galerkin time-domain (DGTD) method based on dynamically adaptive Cartesian meshes (ACM) is developed for a full-wave analysis of electromagnetic fields in dispersive media. Hierarchical Cartesian grids offer simplicity close to that of structured grids and the flexibility of unstructured grids while being highly suited for adaptive mesh refinement (AMR). The developed DGTD-ACM achieves a desired accuracy by refining non-conformal meshes near material interfaces to reduce stair-casing errors without sacrificing the high efficiency afforded with uniform Cartesian meshes. Moreover, DGTD-ACM can dynamically refine the mesh to resolve the local variation of the fields during propagation of electromagnetic pulses. A local time-stepping scheme is adopted to alleviate the constraint on the time-step size due to the stability condition of the explicit time integration. Simulations of electromagnetic wave diffraction over conducting and dielectric cylinders and spheres demonstrate that the proposed method can achieve a good numerical accuracy at a reduced computational cost compared with uniform meshes. For simulations of dispersive media, the auxiliary differential equation (ADE) and recursive convolution (RC) methods are implemented for a local Drude model and tested for a cold plasma slab and a plasmonic rod. With further advances of the charge transport models, the DGTD-ACM method is expected to provide a powerful tool for computations of electromagnetic fields in complex geometries for applications to high-frequency electronic devices, plasmonic THz technologies, as well as laser-induced and microwave plasmas.
\end{abstract}




\end{frontmatter}


\section{Introduction}\label{sec:Intro}
Over the past decades, several numerical methods have been developed to solve Maxwell's equations in the time domain, which include the finite-difference time-domain (FDTD) \cite{bib:FDTD}, the finite-element time-domain (FETD) \cite{bib:Jin:FEM}, and the finite-volume time-domain (FVTD) \cite{bib:Munz:MaxwellFVTD} methods. Despite its widespread use in multiple disciplines due to the simplicity and high efficiency, the FDTD method is severely limited by the structured meshes and low-order spatial and temporal discretizations it employs. The FETD method is very flexible and accurate because of its use of unstructured meshes and higher-order spatial and temporal discretization techniques. However, the use of continuous finite element basis functions \cite{bib:Webb:HierarchalFE, bib:Jiao:Orthogonal-TDFEM} and the implicit time integration schemes \cite{bib:Newmark:Newmark, bib:Gedney:NewmarkFE} in the FETD method result in a global system to solve, which is computationally very intensive for large problems. By introducing the idea of numerical fluxes, the FVTD method decomposes an unstructured mesh into individual elements, the fields in which are evolved locally. Therefore, the FVTD method avoids the solution of a global system, and can achieve high computational and parallel efficiencies. Unfortunately, the FVTD extension to a higher-order spatial resolution can be cumbersome \cite{bib:Hesthaven:HighOrder-TD-Review}.

In \cite{hesthaven2002nodal}, an extension to both the FETD and FVTD methods, called the discontinuous Galerkin time-domain (DGTD) method, was proposed and has shown advantages over the aforementioned time-domain solvers. Similar to the FETD method, the DGTD method is able to employ unstructured meshes, higher-order basis functions, and higher-order time integration methods, resulting in higher-order accuracies in geometrical, spatial, and temporal discretizations. Similar to the FVTD method, the DGTD method adopts the idea of the numerical flux, which makes the computation entirely element-based, and hence the DGTD method is well suited for parallel computing \cite{klockner2009nodal}. These features make the DGTD method a good candidate for simulating electromagnetic (EM) problems with a good accuracy and a high efficiency \cite{bib:Shu:DGDevelopment, bib:Hesthaven-Warburton:NodalDG, bib:Lu:DGTD, bib:Gedney:DGTD, bib:Liu:NodalDGTD2005, bib:Liu:VectorDGTD2010, bib:Liu:EB-DGTD-2015, bib:Jiang:DG-lumport-2013, bib:Descombes:DGTD-EM, bib:Jiang:DGBI-2014, bib:Angulo:DGTD-StateOfArt}.

\begin{figure*}[!t]
  \centering{
  \subfloat[]{\includegraphics[height=1.5in]{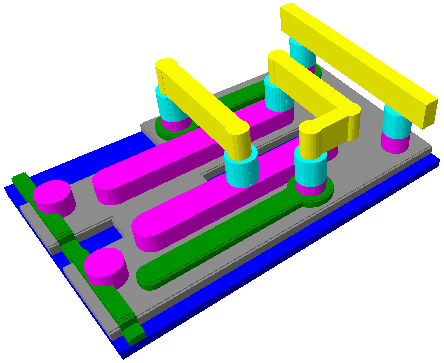}\label{fig:Dev-CAD}} \hfil
  \subfloat[]{\includegraphics[height=1.5in]{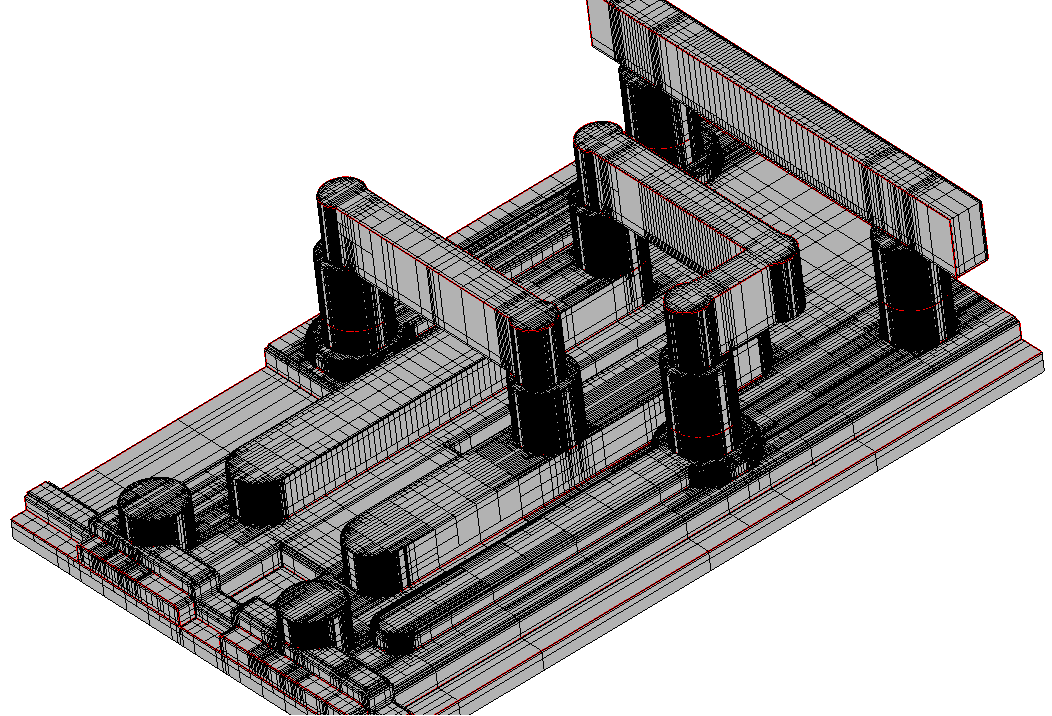}\label{fig:Dev-ACM1}} \hfil
  \subfloat[]{\includegraphics[height=1.5in]{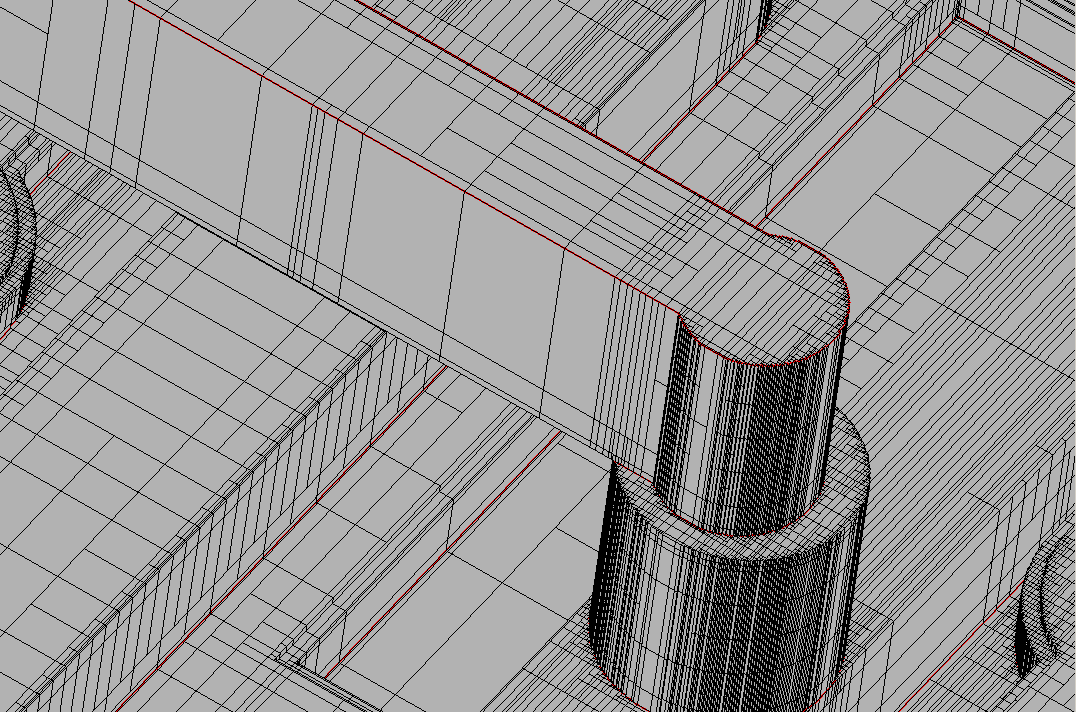}\label{fig:Dev-ACM2}}
  }\caption{An IC setup with electronic interconnects definition (a) and its meshing using an adaptive Cartesian mesh (b) and (c).}
  \label{fig:ACM-Device}
\end{figure*}

Most DGTD algorithms developed so far for computational electromagnetics use boundary-conforming meshes. The computational domain with immersed boundaries or material interfaces is divided into triangular elements in two dimensions and tetrahedral elements in three dimensions. Although these elements can model complex geometries accurately, their use is challenged by the necessity of remeshing for changing geometries. Non-boundary conforming methods avoid remeshing step all together. They are typically implemented for immersed boundaries using structured grids or hierarchical grids. Hierarchical Cartesian grids offer simplicity close to that of structured grids and the flexibility of unstructured grids while being highly suited for adaptive mesh refinement (AMR) as well as for parallelization and dynamic load balancing among processors. Non-boundary conforming methods with AMR have gained popularity in engineering simulations and computer aided design \cite{bib:Weinhold:3rdWaveCFD}. Although these methods start finding their way into computational electromagnetics \cite{bib:Weiland:DGTD-hp-EM}, they are still rarely used. Among notable exceptions are the book \cite{bib:Sarris:AMR-TD-EM}, which describes AMR-FDTD method, and the recent paper \cite{bib:Barbas:EM-AMR}, which uses the Godunov method for solving Maxwell's equations. In this paper, we use an adaptive Cartesian mesh (ACM), which has many attractive features such as automatic mesh generation for complex geometries, dynamic mesh refinement to adapt to the solution and changing geometry, and efficient parallelization \cite{bib:Kolobov:kinetic-fluid}.

The hierarchical ACM is generated by subsequent division of squares (in 2D) or cubes (in 3D), which corresponds to a binary, quad- or octree data structures. In particular, the mesh is adaptively refined near immersed surfaces or material interfaces in order to better resolve the boundary curvature, leading to non-conformal meshes where one hexahedral element interfaces with multiple smaller hexahedral elements. Figure \ref{fig:ACM-Device} shows an example of an integrated circuit (IC) with electronic interconnects meshed using a hierarchical binary mesh, which allows for anisotropic mesh adaptation \cite{bib:CFD-Micromesh}. Automatic mesh adaptation can be applied to device features such as doping profiles, material interfaces and curvatures, to provide a satisfactory geometrical resolution. Such a mesh can also be adapted efficiently when the device geometry changes during an optimization process. This observation motivated the development of a DGTD algorithm for ACM in this work. However, to further enhance the accuracy and efficiency of the DGTD-ACM, there are two more technical issues one has to deal with carefully.

In a numerical simulation, a higher spatial resolution and a better numerical accuracy can be achieved by increasing the mesh density near material interfaces. But when the EM problem is excited by an EM pulse in the time domain, the fields are highly oscillatory around the pulse, but relatively smooth in the rest of the simulation domain. In this case, employing a uniformly dense mesh would increase the number of degrees of freedom (DoFs) significantly. To reduce the computational cost, a dynamic mesh adaptation can be combined with the DGTD method, which changes the resolution of each element at each time step to capture the local variation of the EM fields. However, due to the Courant-Friedrichs-Lewy (CFL) condition, the time-step size for explicit schemes would be constrained by the smallest mesh element. To alleviate this restriction, a local time-stepping (LTS) technique is also adopted with the dynamic mesh adaptation method. In the DGTD method, the computational overhead for reconstructing the mesh and the matrices is proportional to the number of elements changed. When the EM field is smooth in most part of the region, this could achieve a significant reduction of the computational time.

For the modeling of plasmas and electronic devices, computation of electromagnetic fields is only one of the two steps of the complete modeling. The other step is the modeling of particle dynamics, which requires the solution of the charge transport  equations \cite{bib:Klimeck:ComputationalElectronics, bib:Kolobov:adapt-kinetic-fluid}. Since the EM fields provide forces and energy for the particles to evolve and propagate, the EM field components are required to be continuous due to the stability condition for the Boltzmann solver \cite{bib:Assous:Penalty, bib:YAN:Plasma-Shielding}. However, if the DGTD with vector basis functions is employed for the solution of EM fields \cite{bib:Lu:DGTD, bib:Gedney:DGTD, bib:Liu:VectorDGTD2010, bib:Liu:EB-DGTD-2015, bib:Jiang:DG-lumport-2013, bib:Jiang:DGBI-2014}, due to the property of vector basis functions \cite{bib:Nedelec:MixedFE, bib:Webb:HierarchalFE}, only the EM field components that are tangential to the elemental interfaces are continuous, and those that are normal to the interfaces are discontinuous. When coupled with the Boltzmann solver, the discontinuous normal components of the fields will impose an unpredictable and inaccurate amount of force and energy to the particles and result in spurious and erroneous solutions. To overcome this difficulty, the DGTD method with scalar interpolatory (nodal) basis functions \cite{bib:Hesthaven-Warburton:NodalDG, bib:YAN:Plasma-Shielding} is used in this paper to provide a field solution that is continuous in all $x$, $y$, and $z$ directions.

The rest of the paper is organized as follows. In Section \ref{sec:DGTD-Theory}, the theory of the DGTD method is introduced. In Sections \ref{sec:DynamicMeshAdaptation} and \ref{sec:LTS}, the algorithms of dynamic mesh adaptation and LTS are described, respectively. Section \ref{sec:NumericalExamples} presents numerical examples to validate the proposed method and demonstrate the capability of the DGTD solver with ACM and LTS techniques. The modeling methods of dispersive media are presented in Section \ref{sec:Dispersive}, followed by two numerical examples to validate and demonstrate the application of the proposed DGTD-ACM method. The paper is concluded in Section \ref{sec:Conclusion}.

\section{The DGTD Method}\label{sec:DGTD-Theory}
Consider the dynamic Maxwell's equations in a medium characterized by permittivity $\eps$ and permeability $\mu$
\begin{align}
\mu\frac{\partial\v{H}}{\partial t} &= -\nabla\times\v{E}\\
\eps\frac{\partial\v{E}}{\partial t} &= \nabla\times\v{H} -\v{J}
\end{align}
where $\v{E}$ is the electric field, $\v{H}$ is the magnetic field, and $\v{J}$ is the current density. To solve these equations using the DGTD method, the solution domain is first discretized into small elements $K_i$. In this work, structured rectangular and cuboidal elements are employed as the discretization elements in 2D and 3D cases, respectively, due to their simplicity and high efficiency, although other element types can also be used to discretize the solution domain. After the geometric discretization, Maxwell's equations are converted into the strong form by testing with Lagrange polynomials $\v{\phi}_m \!=\! \uv{u} \, \phi_m$ ($\uv{u} \!=\! \uv{x}$, $\uv{y}$, or $\uv{z}$) and applying the Gauss divergence theorem twice in each element to yield
\begin{eqnarray}
\label{eqn:4:3a}
  \frac{\textrm{d}}{\textrm{d}t} \! \int_{K_i} \!
  \v{\phi}_m \! \cdot \! \mu\v{H} \, \textrm{d}V \eq
     - \! \int_{K_i} \! \v{\phi}_m \! \cdot \! \nabla \! \times \! \v{E} \, \textrm{d}V  -\sum_j\int_{S_i^j} \! \v{\phi}_m \! \cdot \! [\uv{n}\!\times\!(\v{E}^{*} \!-\! \v{E})] \, \textrm{d}S \\
\label{eqn:4:3b}
  \frac{\textrm{d}}{\textrm{d}t} \! \int_{K_i} \!
  \v{\phi}_m \! \cdot \! \eps\v{E} \, \textrm{d}V \eq
     \int_{K_i} \! \v{\phi}_m \! \cdot \! \nabla \! \times \! \v{H} \, \textrm{d}V -\int_{K_i} \! \v{\phi}_m \! \cdot \! \v{J} \, \textrm{d}V +\sum_j\int_{S_i^j} \! \v{\phi}_m \! \cdot \! [\uv{n}\!\times\!(\v{H}^{*} \!-\! \v{H})] \, \textrm{d}S
\end{eqnarray}
where $S_i^j$ denotes the $j$-th face of the element $K_i$, $\E^*$ and $\H^*$ denote the intermediate states defined on $S_i^j$. It should be pointed out that in this work, only Faraday's and Amp\`{e}re's laws are considered, while the two Gauss's laws are treated as a natural consequence of the vector identity $\div \nabla \times \!\!=\!\! 0$ and the charge conservation law. In the case of a self-consistent simulation involving particle kinetics, Gauss's laws can be considered and solved with a divergence cleaning technique \cite{bib:Munz:DivCorrection, bib:Munz:divFVTD, bib:Pfeiffer:hdc-pic}.

In the DG formulation, the fields are continuous within each mesh element $K_i$, but are allowed to be discontinuous across the elemental interfaces $S_i^j$. The fields in the two adjacent elements are connected through the intermediate states on each face. As a result, the fields can be solved in each element independently, which results in an element-level domain decomposition scheme. Typical choices of the numerical fluxes include the central and upwind fluxes \cite{bib:Hesthaven-Warburton:NodalDG}. Here the central flux formulation is given for illustration purposes, where the tangential components of the intermediate states on $S_i^j$ are approximated as the average of the fields in two adjacent elements
\begin{eqnarray}
  \label{eqn:4:5a}
  \n \cross \E^* \eq \frac{1}{2} \ \n \cross \fun{\E^+ + \E^-}
  \\
  \label{eqn:4:5b}
  \n \cross \H^* \eq \frac{1}{2} \ \n \cross \fun{\H^+ + \H^-}
\end{eqnarray}
in which $\v{E}^{-}$ and $\v{H}^{-}$ denote the fields in element $K_i$, $\v{E}^{+}$ and $\v{H}^{+}$ denote the fields in its neighboring element, and the unit normal vector $\n$ points from element $K_i$ to its neighbor.

Expanding the unknown fields with the $p$-th order Lagrange polynomials $\phi_n$ as
\begin{eqnarray}
  \label{eqn:4:7a}
  \hspace{-3pt}\E \fun{\v{r},t} \!\eq\! \sum^{N_p-1}_{n=0} \phi_n \!\fun{\v{r}} \mat{ \uv{x} E_{xn}\!\fun{t} + \uv{y} E_{yn}\!\fun{t} + \uv{z} E_{zn}\!\fun{t} } \\
  \label{eqn:4:7b}
  \hspace{-3pt}\H \fun{\v{r},t} \!\eq\! \sum^{N_p-1}_{n=0} \phi_n \!\fun{\v{r}} \mat{ \uv{x} H_{xn}\!\fun{t} + \uv{y} H_{yn}\!\fun{t} + \uv{z} H_{zn}\!\fun{t} }
\end{eqnarray}
where $N_p=\fun{p+1}^d$ stands for the number of DoFs in a $d$-dimensional element, the strong form of Maxwell's equations can be converted into the matrix form representation as
\begin{eqnarray}
  \label{eqn:4:8a}
  \mat{M} \d{}{t}\vec{H_x} \eq -\frac{1}{\mu} \big( \mat{S_y} \vec{E_z} - \mat{S_z} \vec{E_y} + \mat{M_{\textrm{f}}} \vec{F^E_x} \big) \\
  \label{eqn:4:8b}
  \mat{M} \d{}{t}\vec{H_y} \eq -\frac{1}{\mu} \big( \mat{S_z} \vec{E_x} - \mat{S_x} \vec{E_z} + \mat{M_{\textrm{f}}} \vec{F^E_y} \big) \\
  \label{eqn:4:8c}
  \mat{M} \d{}{t}\vec{H_z} \eq -\frac{1}{\mu} \big( \mat{S_x} \vec{E_y} - \mat{S_y} \vec{E_x} + \mat{M_{\textrm{f}}} \vec{F^E_z} \big) \\
  \label{eqn:4:8d}
  \mat{M} \d{}{t}\vec{E_x} \eq \frac{1}{\eps} \big( \mat{S_y} \vec{H_z} - \mat{S_z} \vec{H_y} + \mat{M_{\textrm{f}}} \vec{F^H_x} - \mat{M} \vec{J_x} \big) \\
  \label{eqn:4:8e}
  \mat{M} \d{}{t}\vec{E_y} \eq \frac{1}{\eps} \big( \mat{S_z} \vec{H_x} - \mat{S_x} \vec{H_z} + \mat{M_{\textrm{f}}} \vec{F^H_y} - \mat{M} \vec{J_y} \big) \\
  \label{eqn:4:8f}
  \mat{M} \d{}{t}\vec{E_z} \eq \frac{1}{\eps} \big( \mat{S_x} \vec{H_y} - \mat{S_y} \vec{H_x} + \mat{M_{\textrm{f}}} \vec{F^H_z} - \mat{M} \vec{J_z} \big)
\end{eqnarray}
where the elements of the mass, stiffness, and facial mass matrices are given by ($u\!=\!x$, $y$, or $z$)
\begin{eqnarray}
  \label{eqn:4:11a}
  \mat{M}_{mn}   \eq \int_{K_i}{\phi_m \, \phi_n \, \textrm{d}V} \\
  \label{eqn:4:11b}
  \mat{S_u}_{mn} \eq \int_{K_i}{\phi_m \, \frac{\partial \phi_n}{\partial u} \, \textrm{d}V} \\
  \label{eqn:4:11c}
  \mat{M_{\textrm{f}}}_{mn} \eq \int_{S_i^j}{\phi_m \, \phi_n \, \textrm{d}S}
\end{eqnarray}
and the numerical fluxes are given by
\begin{eqnarray}
  \label{eqn:4:9}
  \v{F}^E \eq \frac{1}{2} \n \cross \fun{\E^+ - \E^-} \\
  \v{F}^H \eq \frac{1}{2} \n \cross \fun{\H^+ - \H^-}.
\end{eqnarray}

Once the time derivatives ${\textrm{d}}{\vec{E_u}}/{\textrm{d}}{t}$ and ${\textrm{d}}{\vec{H_u}}/{\textrm{d}}{t}$ are obtained, the classic fourth-order Runge-Kutta method \cite{bib:Butcher:NumericalODE} can be applied to (\ref{eqn:4:8a})-(\ref{eqn:4:8f}) to advance the EM fields from $t_n$ to $t_{n+1} = t_n + \dt$ as
\begin{eqnarray}
\v{q}\fun{t_{n+1}} \eq \v{q}\fun{t_n} + \frac{\dt}{6} \fun{\v{k}_1 + 2 \v{k}_2 + 2 \v{k}_3 + \v{k}_4}
\end{eqnarray}
in which
\begin{eqnarray}
\v{k}_1 \eq f \mat{ t_n, \v{q}\fun{t_n} }\label{eq:RK4-k1}\\
\v{k}_2 \eq f \! \mat{ t_n\!+\!\frac{\dt}{2}, \v{q}\fun{t_n} + \frac{\dt}{2}\v{k}_1 }\\
\v{k}_3 \eq f \! \mat{ t_n\!+\!\frac{\dt}{2}, \v{q}\fun{t_n} + \frac{\dt}{2}\v{k}_2 }\\
\v{k}_4 \eq f \mat{ t_n+\dt,                \v{q}\fun{t_n} + \dt \v{k}_3               }\label{eq:RK4-k4}
\end{eqnarray}
where $\v{q}(t) \!=\! \mat{\vec{E_x}, \vec{E_y}, \vec{E_z}, \vec{H_x}, \vec{H_y}, \vec{H_z}}^{\textrm{T}}$ denotes the EM field vector, $f\mat{t, \v{q}(t)} = \textrm{d} \v{q} / \textrm{d}t$ stands for the operations in order to obtain the time derivatives, and $\v{k}_i$ ($i=1,2,3,4$) are known as the stage vectors. To maintain the stability of the time integration scheme, the time-step size $\dt$ is limited by the CFL condition \cite{bib:Shu:RKDG-Convection, leveque2002finite}
\begin{eqnarray}
\dt \leq \frac{1}{2p+1} \frac{h}{c}
\end{eqnarray}
where $c$ denotes the speed of light, $p$ denotes the basis order, and $h$ denotes the size of the element $K_i$. If a uniform time-step size is used, the minimum element size throughout the simulation domain should be used as $h$.

\begin{figure}[!t]
  \centering
  \includegraphics[scale = 0.5]{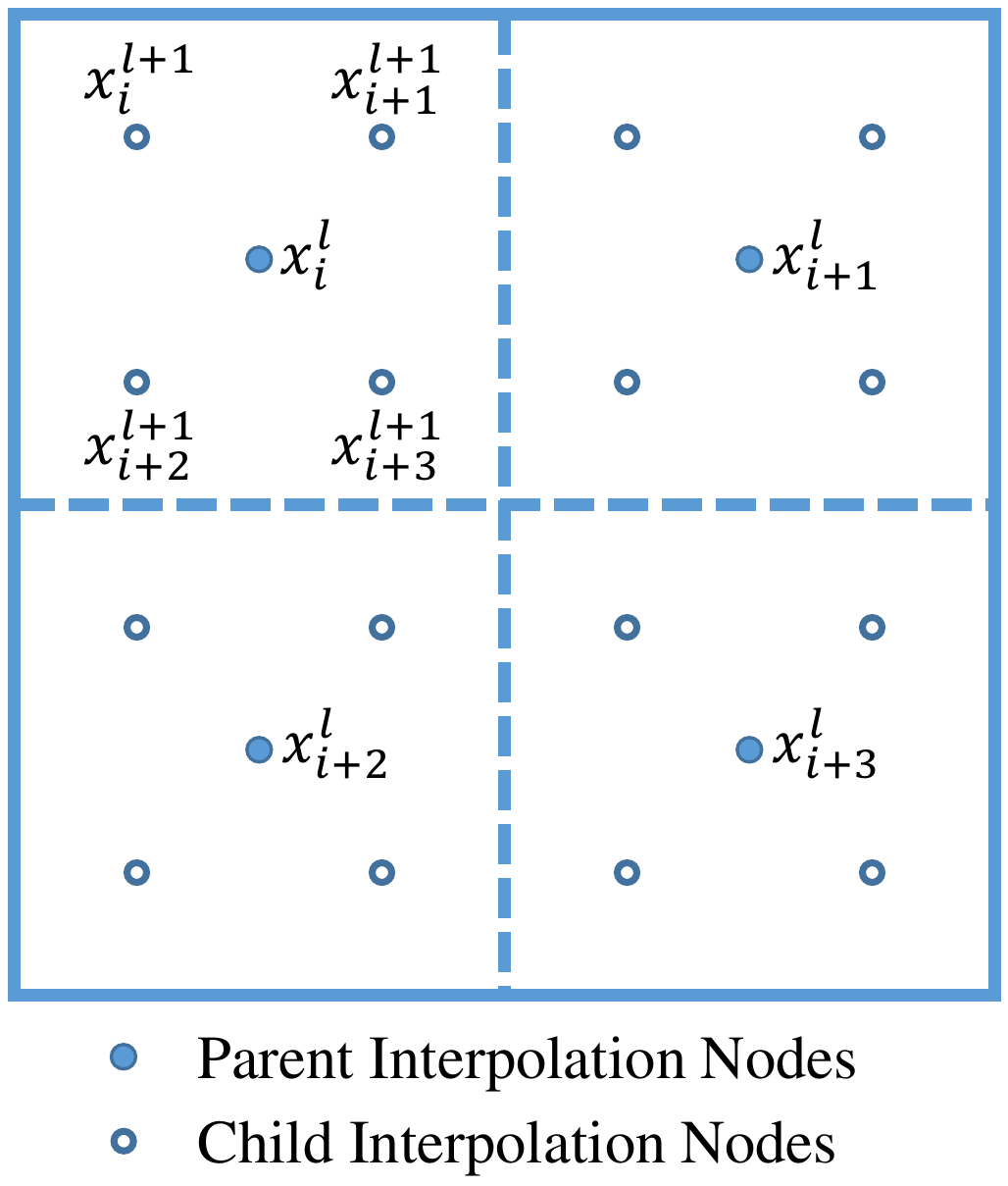}
  \caption{Illustration of the coarse-to-fine and fine-to-coarse element mapping during the mesh adaptation in 2D.}
  \label{fig:DACM}
\end{figure}

\section{Dynamic $h$-Adaptation Technique}\label{sec:DynamicMeshAdaptation}
To resolve the fields with drastic spatial variations, finer grids are needed. In simulations of pulse propagation and scattering problems, a direct application of a uniformly fine mesh throughout the solution domain would increase the computational cost significantly. Whenever the mesh is refined by $s$ times, the computational cost will increase by a factor of $s^{d+1}$ in a $d$-dimensional problem, where the extra one on the exponent comes from the CFL condition imposed by the explicit time integration method. In order to achieve a good spatial resolution without significantly increasing the computational cost, a dynamic mesh adaptation technique based on the ACM \cite{popinet2003gerris} is developed in this work to enhance the DGTD algorithm, where the sizes of the cells are dynamically adjusted according to the variation of the local fields. To implement the dynamic mesh adaptation, three issues need to be addressed. The first one is the criteria of the mesh refinement and coarsening. The second issue is the approach to reconstructing the DoFs after a mesh has changed. The last issue is how to connect mesh elements with different cell sizes. Each of these issues is addressed here as follows.

\subsection{Dynamic Mesh Refinement}
At the beginning of the numerical simulation, a base mesh is first constructed, which should have basic resolution to the wave physics. This base mesh is named as the level-$0$ mesh, and its individual mesh element is named as the level-$0$ element $K_i^0$. The subdivision of a level-$l$ element $K_i^l$ generates $2^d$ smaller elements $K_i^{l+1}$ on level $l+1$.

As the fields start to propagate into the solution domain and impinge on the geometry, their distribution become nonuniform in the domain. To determine the correct element size at a given location and time, the variation of the EM fields can be measured through different means, such as the gradient of the local EM power density $\nabla \fun{\eps \|\E\|^2 + \mu \|\H\|^2}$, or the local electric field intensity $\nabla E_v$  ($v = x, y, z$) at the interpolation nodes $\v{x}_{i+n}^l$ ($n=1,\ldots,N_p$) in element $K_i^l$.  For example, element $K_i^l$ is refined if the $L_2$ norm of the gradient of the electric field component satisfies one of the following criteria
\begin{eqnarray}
\left\| \nabla E_v \right\| > \zeta_{\textrm{max}} \max \vec{\left\| \nabla E_v \right\|}
\end{eqnarray}
or
\begin{eqnarray}
\left\| \nabla E_v \right\| > \xi_{\textrm{max}}
\end{eqnarray}
where $\zeta_{\textrm{max}}$ and $\xi_{\textrm{max}}$ are preset thresholds, and $\max \vec{\left\| \nabla E_v \right\|}$ is the maximum gradient value throughout the entire simulation domain. Element $K_i^l$ and its neighbouring elements are coarsened to a larger element at level $(l-1)$ if both of the following criteria are satisfied
\begin{eqnarray}
\left\| \nabla E_v \right\| &<& \zeta_{\textrm{min}} \max \vec{\left\| \nabla E_v \right\|} \\
\left\| \nabla E_v \right\| &<& \xi_{\textrm{min}}
\end{eqnarray}
where $\zeta_{\textrm{min}}$ and $\xi_{\textrm{min}}$ are the preset thresholds.

In the DGTD method, the partial derivatives in the gradient operation
\begin{eqnarray}
\nabla E_v = \sum_u \uv{u} \pd{E_v}{u}
\end{eqnarray}
can be easily obtained by matrix-vector product between the stiffness matrix $\mat{S_u}$ and the field component vector $\vec{E_v}$ as
\begin{eqnarray}
\pd{E_v}{u} = \mat{S_u} \vec{E_v}.
\end{eqnarray}

\subsection{DoF Reconstruction}
During dynamic cell/element coarsening and refinement, larger and smaller cells/elements are created on-the-fly. This requires the solutions being mapped from fine to coarse cells and from coarse to fine cells. The particularity of the DGTD technique is that the electric and magnetic fields are defined on several interpolation nodes in the cells instead of only the cell centers as in the FVTD codes. For the 4-element (in 2D) and 8-element (in 3D) DG scheme, the quad/octree topology matches the element topology, and the interpolations can be performed in a straightforward and numerically efficient manner. Figure \ref{fig:DACM} is an illustration of such a process, where the solid and empty circles represent the interpolation nodes of the first-order basis functions in the coarse and fine mesh elements, respectively.

The DoFs at $\v{x}_{i+k}^{l+1}$ ($k = 0, 1, 2, 3$) in a fine element on level $l+1$ can be directly obtained from the coarse element $K_i^l$ on level $l$ by interpolating the DoFs at $\v{x}_{i+k'}^{l}$ using their respective basis functions $\phi_{k'}$ as
\begin{align}
E_v \left( \v{x}_{i+k}^{l+1} \right) &= \sum_{k'=0}^{3} E_v \! \left( \v{x}_{i+k'}^{l} \right) \phi_{k'}\! \left( \v{x}_{i+k}^{l+1} \right) \\
H_v \left( \v{x}_{i+k}^{l+1} \right) &= \sum_{k'=0}^{3} H_v\! \left( \v{x}_{i+k'}^{l} \right) \phi_{k'}\! \left( \v{x}_{i+k}^{l+1} \right).
\end{align}
To obtain the DoFs at $\v{x}_{i+k'}^l$ ($k' = 0, 1, 2, 3$) in a coarse element on level $l$ from the fine elements $K_{i+k}^{l+1}$, a similar interpolation formula can be used. If first-order basis functions are employed, the EM fields on each interpolation node of the coarse element are simply the average of the DoFs of the corresponding fine elements. In this case
\begin{align}
E_v(x_{i}^{l}) &= \frac{1}{4}\sum_{k = 0}^3 E_v(x_{i+k}^{l+1})\\
H_v(x_{i}^{l}) &= \frac{1}{4}\sum_{k = 0}^3 H_v(x_{i+k}^{l+1}).
\end{align}

\subsection{Numerical Flux Calculation}
After the DoFs are reconstructed in the newly refined mesh, the DGTD method can be used to advance the DoFs in time using (\ref{eqn:4:3a}) and (\ref{eqn:4:3b}). Compared with the case of a uniform mesh, all the volume integrals remain the same in the case of the adaptive mesh. Since DGTD is an element-level domain decomposition method, only the DoFs within each element are needed in the volume integrals. The only terms that need modification are the surface integral terms, because they need the information from their adjacent elements. When one element interfaces with $q$ adjacent smaller elements, the corresponding surface integral is broken into $q$ surface integrals with a smaller support. For example, if the central flux is used, the surface integral involving electric fields becomes
\begin{eqnarray}
\label{eqn:4:12}
\int_{S_i^j} \! \v{\phi}_m \! \cdot \! [\uv{n}\!\times\!(\v{E}^{*} \!-\! \v{E})] \, \textrm{d}S
\eq \frac{1}{2} \int_{S_i^j} \! \v{\phi}_m \! \cdot \! [\uv{n}\!\times\!(\v{E}^+ \!-\! \v{E}^-)] \, \textrm{d}S \nonumber \\
\eq \frac{1}{2} \! \mat{ \sum_{k=1}^q \int_{S^i_{jk}} \!\!\! \v{\phi}_m \! \cdot \! \uv{n}\!\times\!\v{E}^+ \, \textrm{d}S - \int_{S^i_{j}} \!\! \v{\phi}_m \! \cdot \! \uv{n}\!\times\!\v{E}^- \, \textrm{d}S }\!.
\end{eqnarray}
On each small domain of integration, the Gauss quadrature rule can be applied to perform the surface integrals. Once the numerical fluxes are obtained, the neighbouring elements can be connected, and the entire system can be advanced for one time step $\dt$.

\section{Local Time-Stepping Scheme}\label{sec:LTS}
For the time integration, the explicit Runge-Kutta methods are usually used. One very common choice is the classic four-stage fourth-order Runge-Kutta method as given in (\ref{eq:RK4-k1})--(\ref{eq:RK4-k4}). Despite their simplicities, the biggest drawback of the explicit methods is the constraint on the time-step size in order to maintain stability. According to the CFL condition \cite{bib:Shu:RKDG-Convection, leveque2002finite}, the time-step size is limited by the element with the smallest size. When dynamic meshes with varying elemental sizes are used, the small size of a refined mesh element would result in a very small time-step size, which will greatly reduce the overall efficiency of the simulation. To alleviate the restriction coming from the smallest mesh element, a non-uniform time-step Runge-Kutta scheme \cite{liu2010nonuniform} is adopted, which allows each element to advance in time with its own time-step size, and is known as the LTS method.

\begin{figure}[!t]
  \centering
  \includegraphics[scale = 0.4]{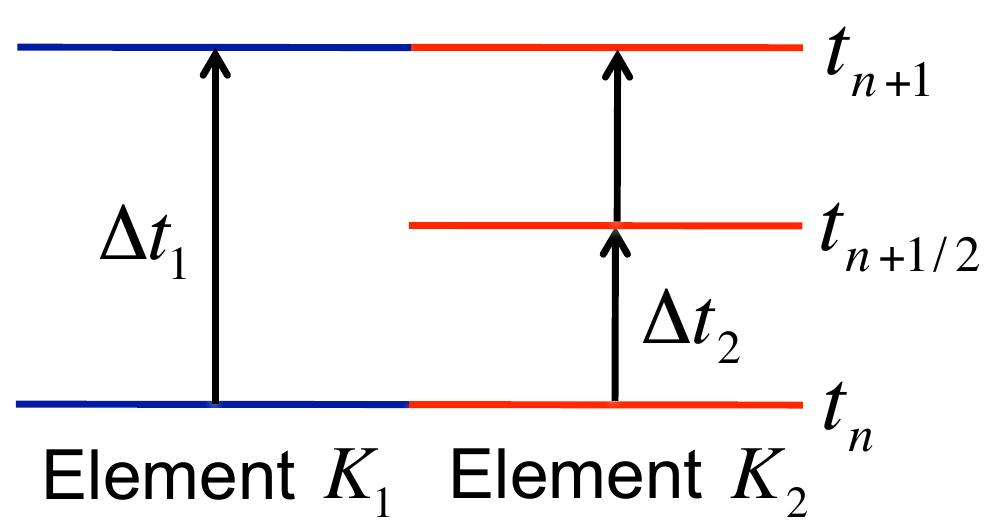}
  \caption{Illustration of the local time-stepping method.}
  \label{fig:LTS}
\end{figure}

The basic idea and formulation of the LTS method are given in this section. Shown in Fig. \ref{fig:LTS} is an illustration of the scheme, where the CFL condition imposes a time-step size $\dt_1$ for element $K_1$ and a time-step size $\dt_2=\dt_1/2$ for element $K_2$. To advance EM fields in elements $K_1$ and $K_2$ using their respective time-step sizes, two scenarios need to be taken into consideration. One is at the synchronized step $t_n$ where the fields in both elements advance simultaneously. The other is the intermediate step $t_{n+1/2}$ where only the fields in element $K_2$ advance in time.

\subsection{Synchronized Step}
At the synchronized step, the stage vectors $\vec{\v{k}_l}$ in (\ref{eq:RK4-k1})--(\ref{eq:RK4-k4}) need to be evaluated at $t=t_n$ in both elements $K_1$ and $K_2$ using time-step size $\dt_1$ and $\dt_2$, respectively. In a linear problem, the stage vectors needed to advance for $\dt_i$ can be related to the time derivatives at $t_n$ as
\begin{align}
  \hspace{-10pt}
  \begin{bmatrix}
  \v{k}_1(\dt_i) \\
  \v{k}_2(\dt_i) \\
  \v{k}_3(\dt_i) \\
  \v{k}_4(\dt_i)
  \end{bmatrix}
  \!\! = \!\!
  \begin{bmatrix}
  1                     \\
  1 & 0.5             \\
  1 & 0.5 &  0.25  \\
  1 &  1   &  0.5  & 0.25
  \end{bmatrix} \!\!\!
  \begin{bmatrix}
  1                     \!  \\
  &     \dt_i    \!         \\
  & &   \!\!\! \dt_i^2   \\
  & & & \!\!\!\!\dt_i^3
  \end{bmatrix} \!\!\!
  \begin{bmatrix}
  \v{q}^{(1)}(t_n) \\
  \v{q}^{(2)}(t_n) \\
  \v{q}^{(3)}(t_n) \\
  \v{q}^{(4)}(t_n)
  \end{bmatrix} \!\!\!
\label{eq:LTS}
\end{align}
where $\v{q}^{(j)}$ stands for the $j$-th order time derivative.

Using (\ref{eq:LTS}), each vector, $\v{k}_{l}(\dt_1)$ in $K_2$, can be inferred from the set $\{\v{k}_{l'}(\dt_2)\ |\ l' = 1, \cdots, l\}$ in $K_2$, and the vector $\v{k}_{l+1}(\dt_1)$ in $K_1$ can be evaluated through (\ref{eq:RK4-k1})--(\ref{eq:RK4-k4}). The same idea applies to the evaluation of $\{\v{k}_{l+1}(\dt_2)\}$ in $K_2$ and therefore $\{\v{k}_{l}(\dt_1)\}$ in $K_1$ and $\{\v{k}_{l}(\dt_2)\}$ in $K_2$ can be obtained simultaneously at $t = t_n$.

\subsection{Intermediate Step}
To update the EM fields in element $K_2$ at $t=t_{n+1/2}$, the vector $\{\v{k}_{l}(\dt_2)\}$ need to be calculated at $t=t_{n+1/2}$. This can be obtained through (\ref{eq:RK4-k1})--(\ref{eq:RK4-k4}) by extrapolating the EM fields and their time derivatives, $\{\v{q}^{(k)}\}$, at $t = t_{n+1/2}$ in element $K_1$ and exploiting (\ref{eq:LTS}) to obtain the vectors $\{\v{k}_{l}(\dt_2)\}$ in $K_1$ at $t = t_{n+1/2}$. To perform the extrapolation, the time derivatives of the EM fields, $\{\v{q}^{(k)}(t_n)\}$ in $K_1$, are obtained through applying (\ref{eq:LTS}) to the vectors $\{\v{k}_l(\dt_1)\}$ in $K_1$ at $t = t_n$. The EM fields and the time derivatives, $\{\v{q}^{(k)}(t_{n+1/2})\}$ in $K_1$, are then estimated through Taylor expansion
\begin{align}
\v{q}^{(k)}(t_{n+1/2}) =
  \sum_{l = k}^{4}\frac{\dt_2^{l-k}}{(l-k)!}\v{q}^{(k)}(t_{n})\text{,  }
    k = 0, \ldots, 4
\end{align}
and the vectors $\{\v{k}_{l}(\dt_2)\}$ at $t = t_{n+1/2}$ can be evaluated by applying (\ref{eq:LTS}) to $\{\v{q}^{(k)}(t_{n+1/2})\}$ again.

\begin{figure}[!t]
  \centering{
  \subfloat[]{\includegraphics[width=2.1in]{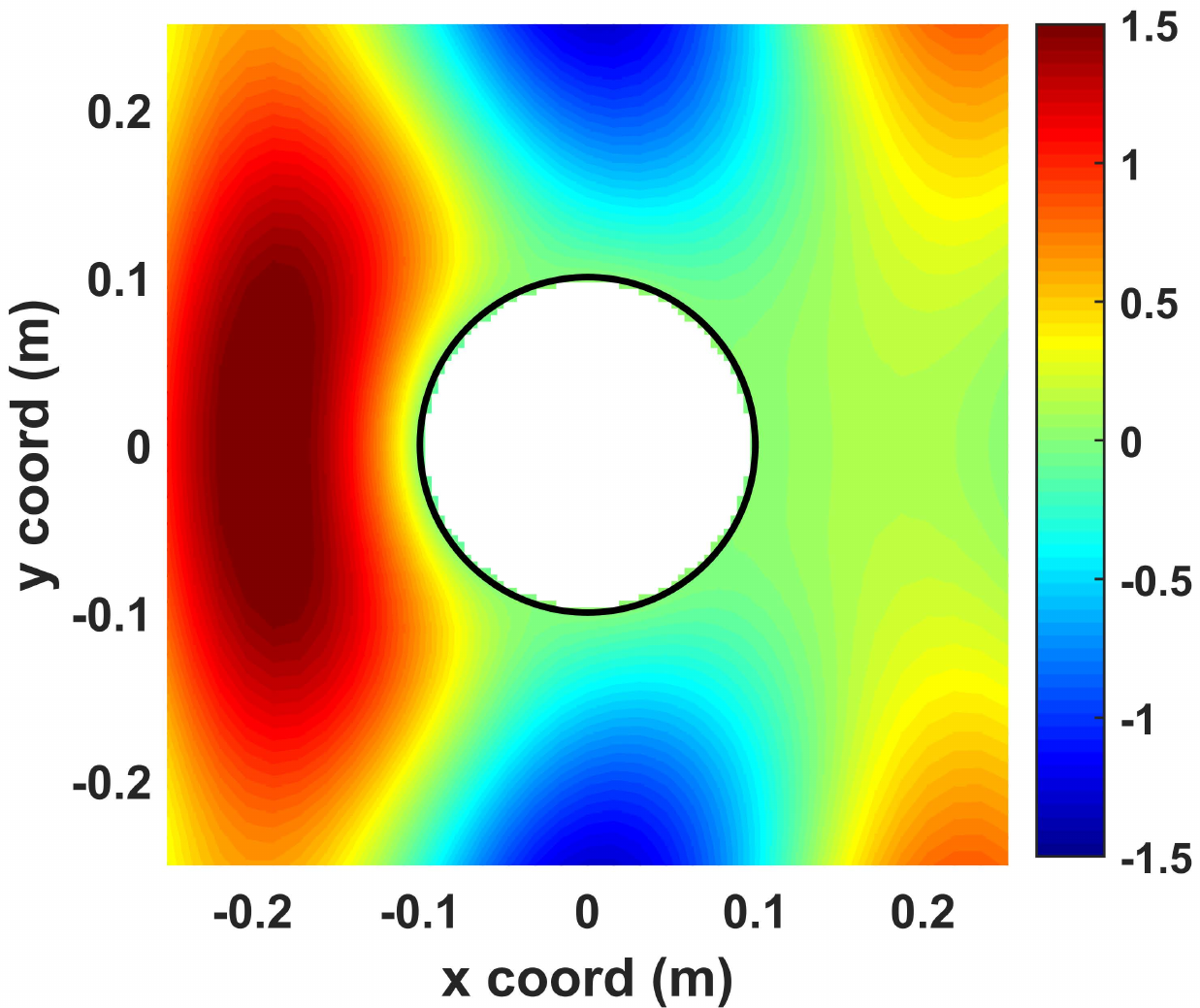}
  \label{fig:CR-1}} \hfil
  \subfloat[]{\includegraphics[width=2.1in]{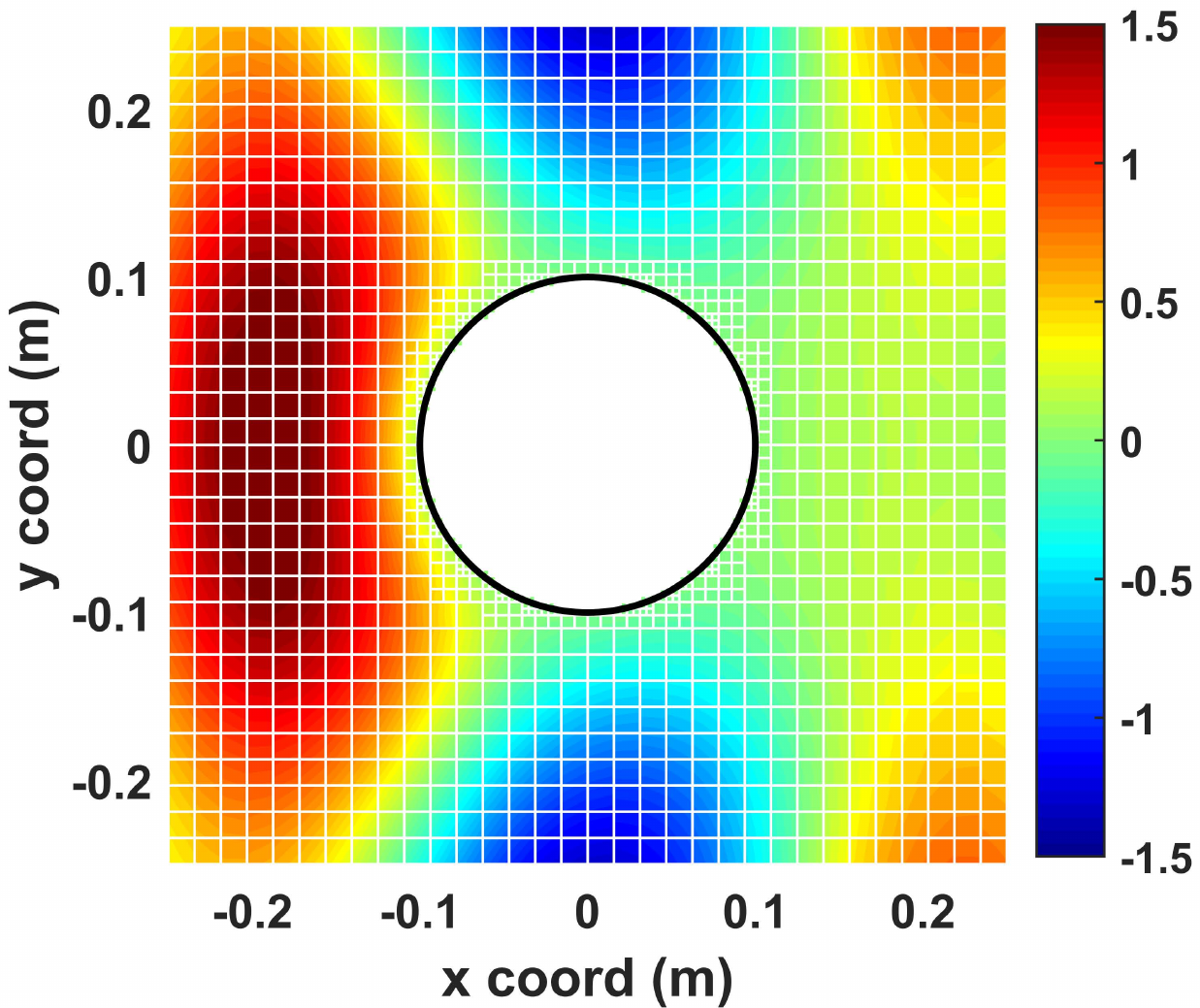}
  \label{fig:CR-2}} \hfil
  \subfloat[]{\includegraphics[width=2.1in]{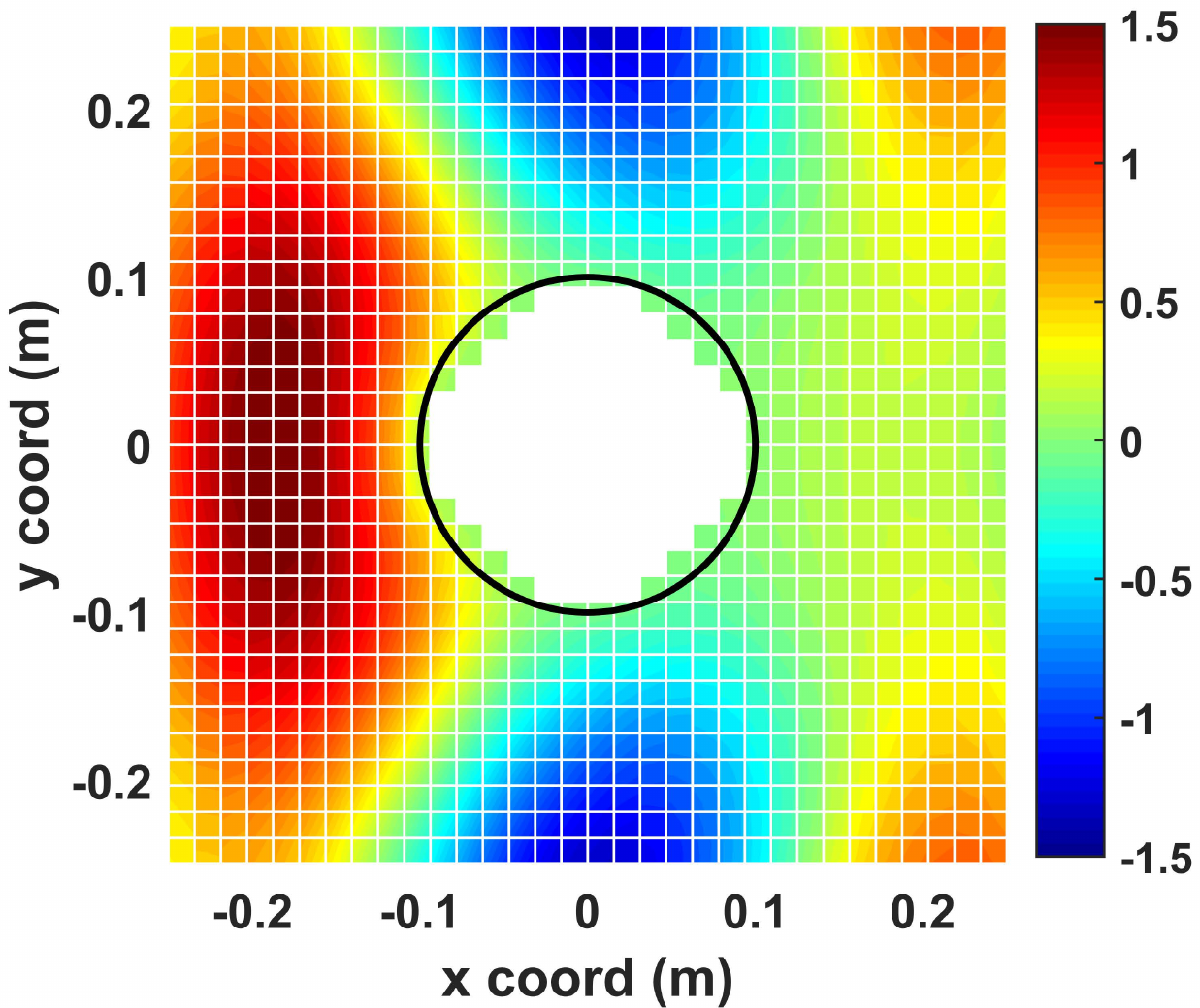}
  \label{fig:CR-3}}
  }\caption{Electric field distributions in the conducting cylinder case at $14.0$ ns. (a) Analytical result; (b) Numerical result obtained using the ACM grid $h_0+2$; and (c) Numerical result obtained using the uniformly coarse grid $h_0+0$.}
  \label{fig:ConductingRod}
\end{figure}

\begin{figure}[!t]
  \centering{
  \subfloat[]{\includegraphics[width=2.1in]{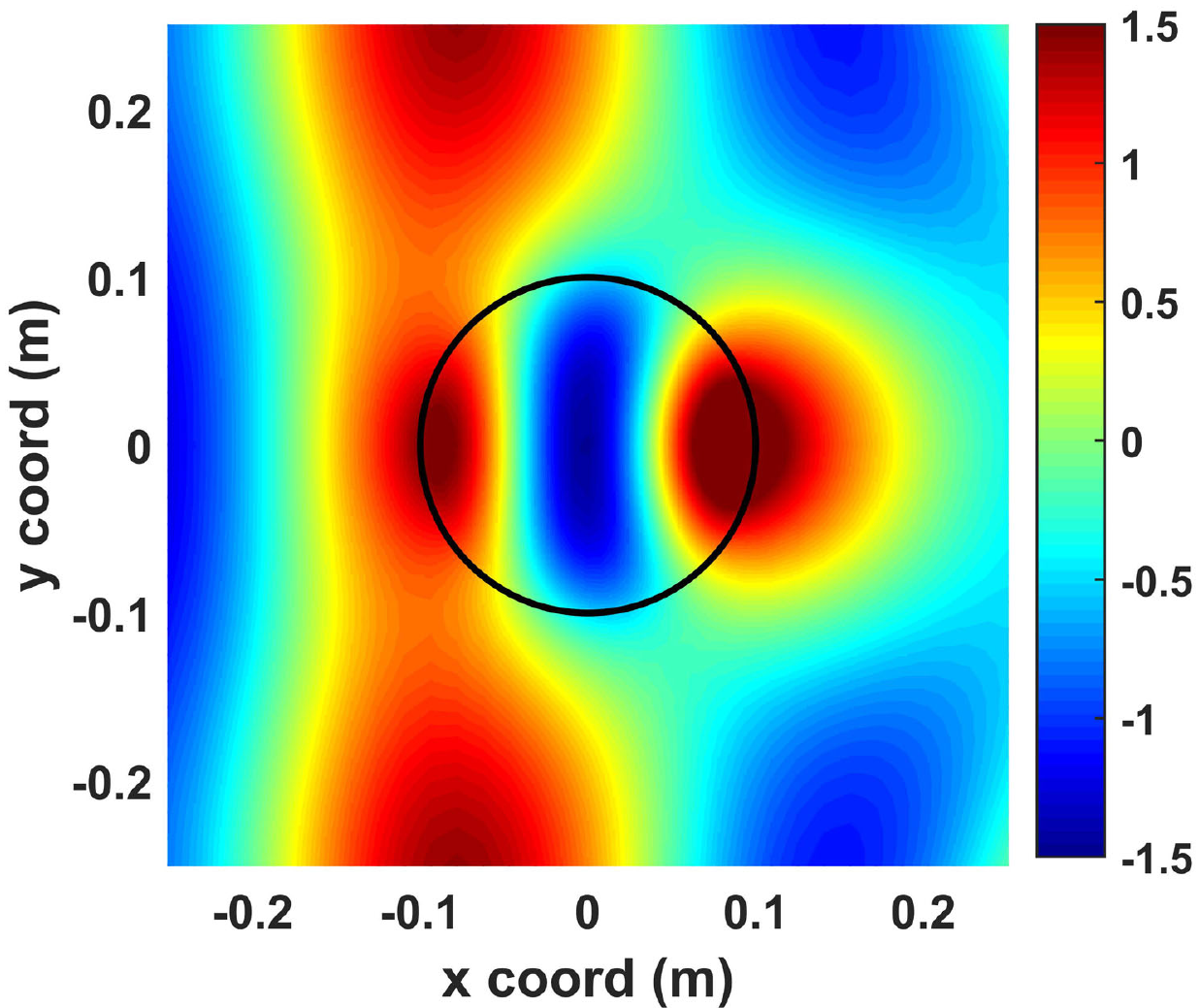}
  \label{fig:ZI-DR-1}} \hfil
  \subfloat[]{\includegraphics[width=2.1in]{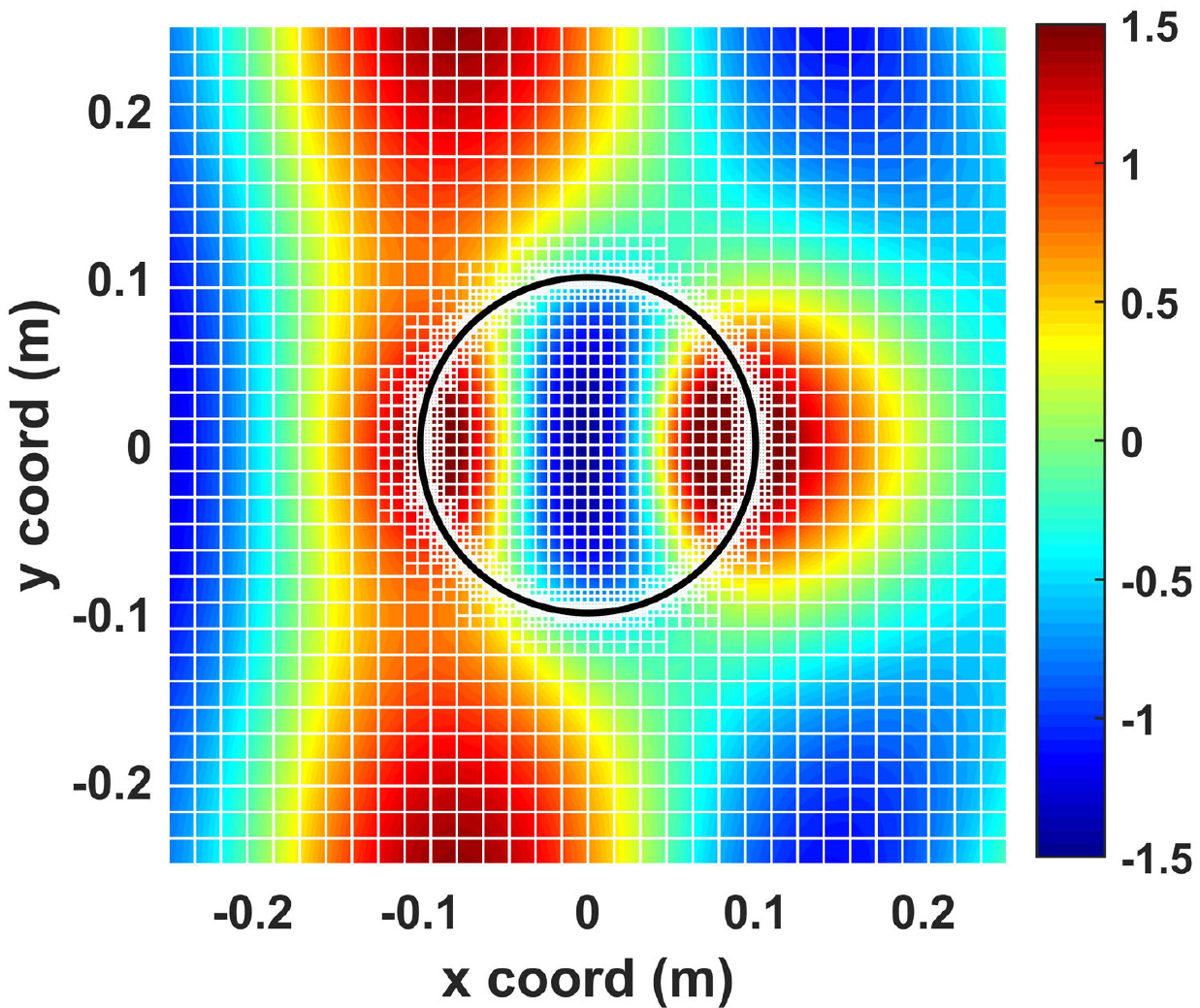}
  \label{fig:ZI-DR-2}} \hfil
  \subfloat[]{\includegraphics[width=2.1in]{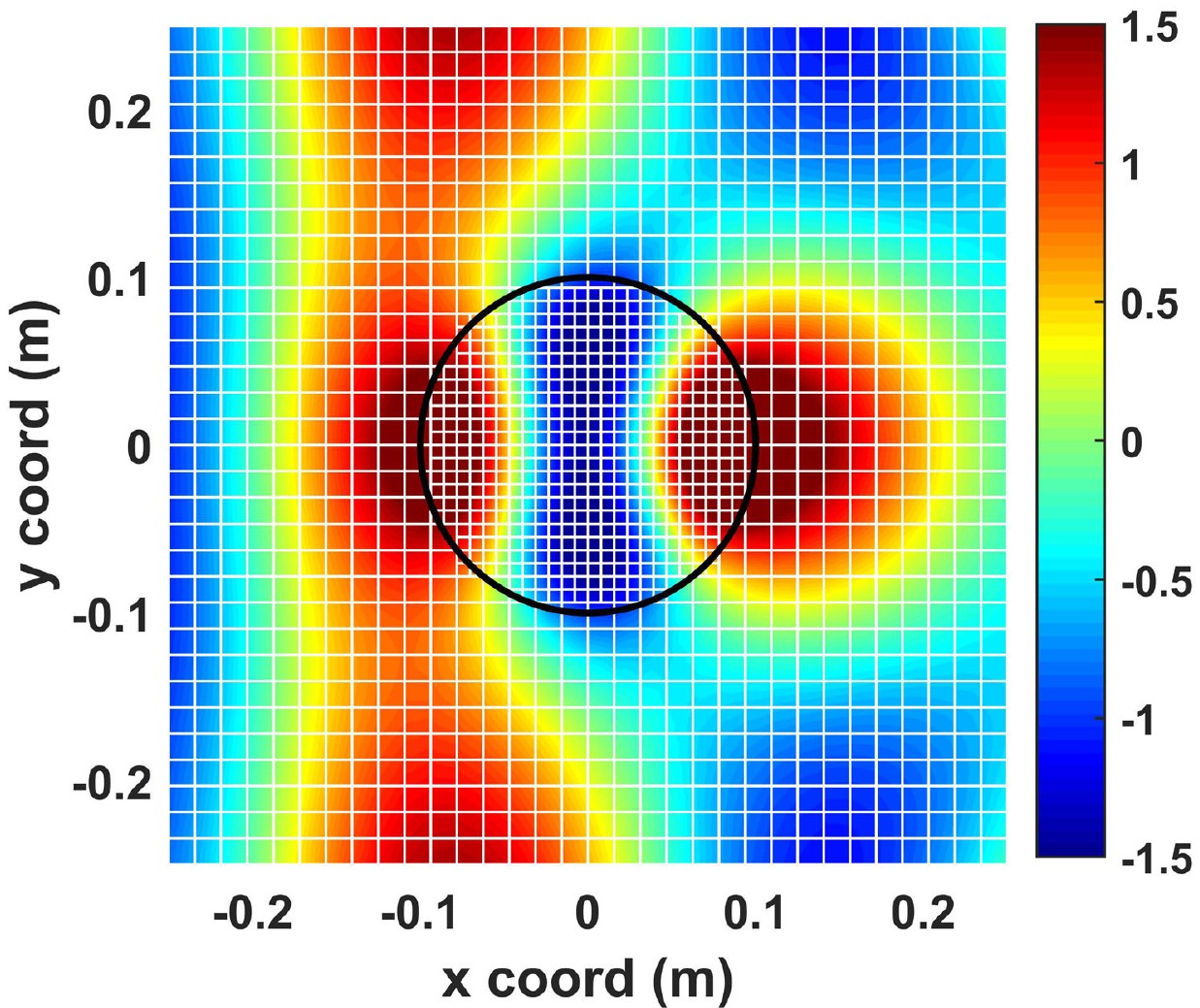}
  \label{fig:ZI-DR-3}}
  }\caption{Electric field distributions in the dielectric cylinder case at $17.1$ ns. (a) Analytical result; (b) Numerical result obtained using the ACM grid $h_0+3$; and (c) Numerical result obtained using the ACM grid $h_0+1$.}
  \label{fig:ZIDielectrocRod}
\end{figure}

\begin{figure}[!t]
  \centering{
  \subfloat[]{\raisebox{20pt}{\includegraphics[width=2.1in]{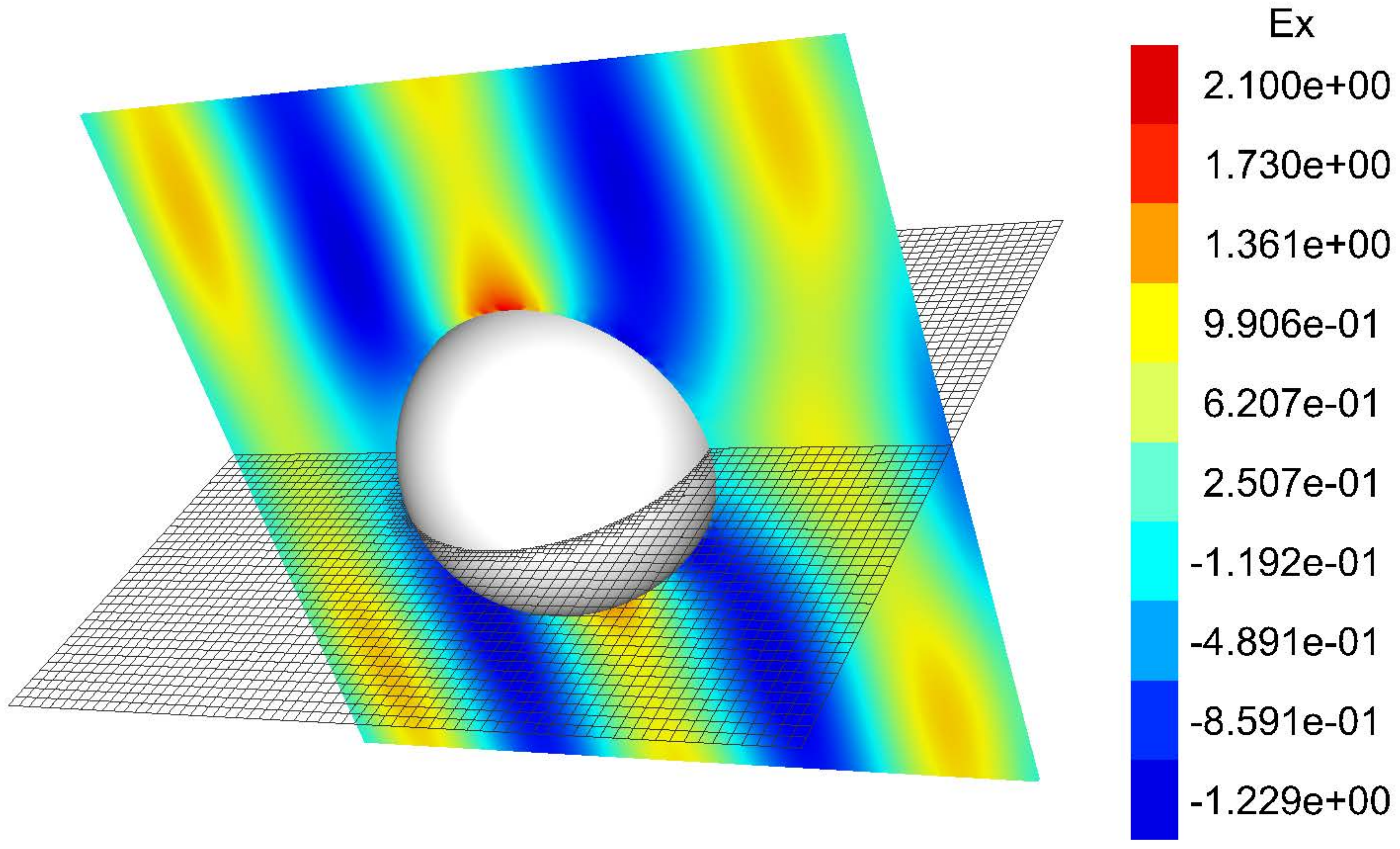}\label{fig:Sphere-2D}}} \hfil
  \subfloat[]{\includegraphics[width=2.1in]{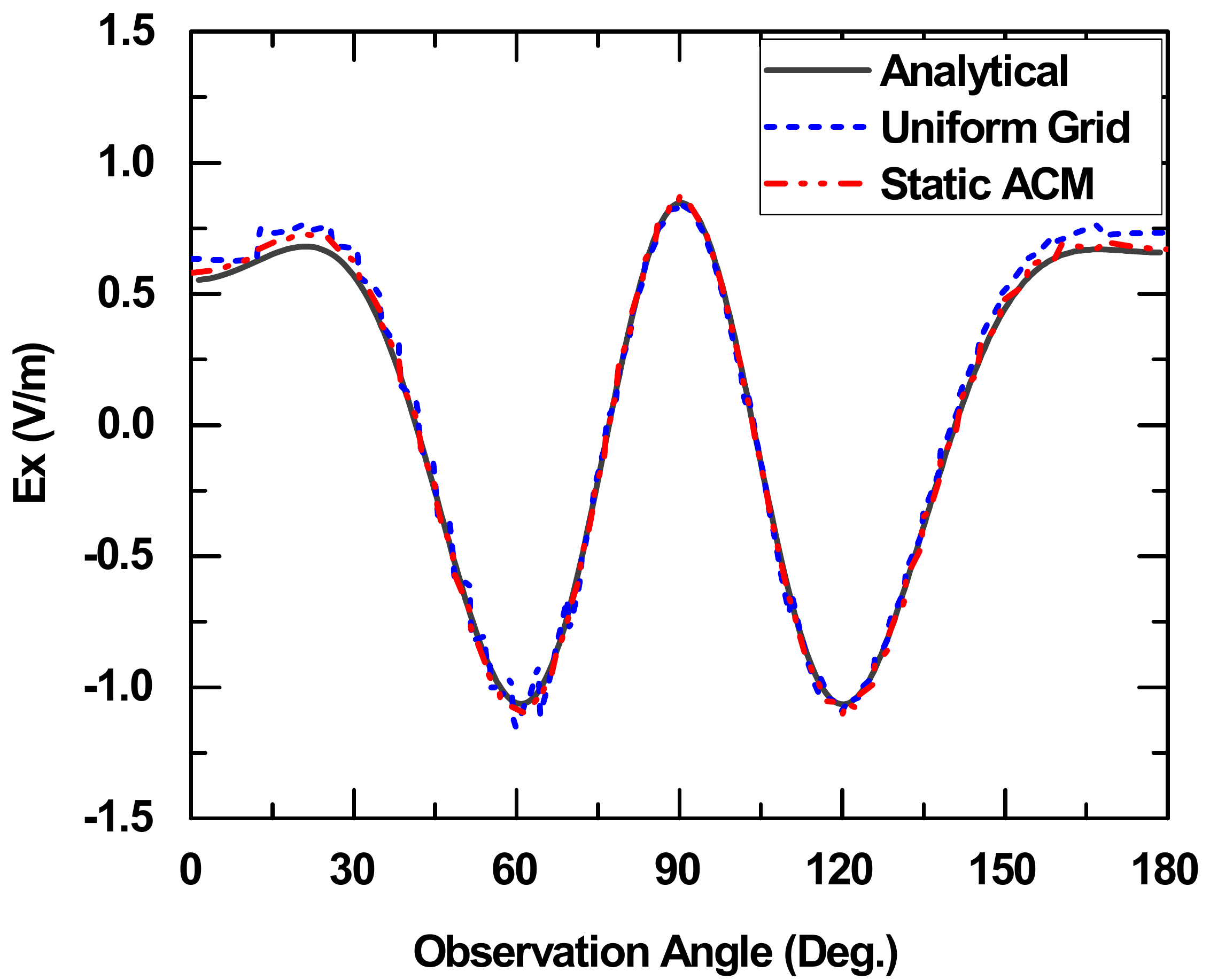}\label{fig:Sphere-xz}} \hfil
  \subfloat[]{\includegraphics[width=2.1in]{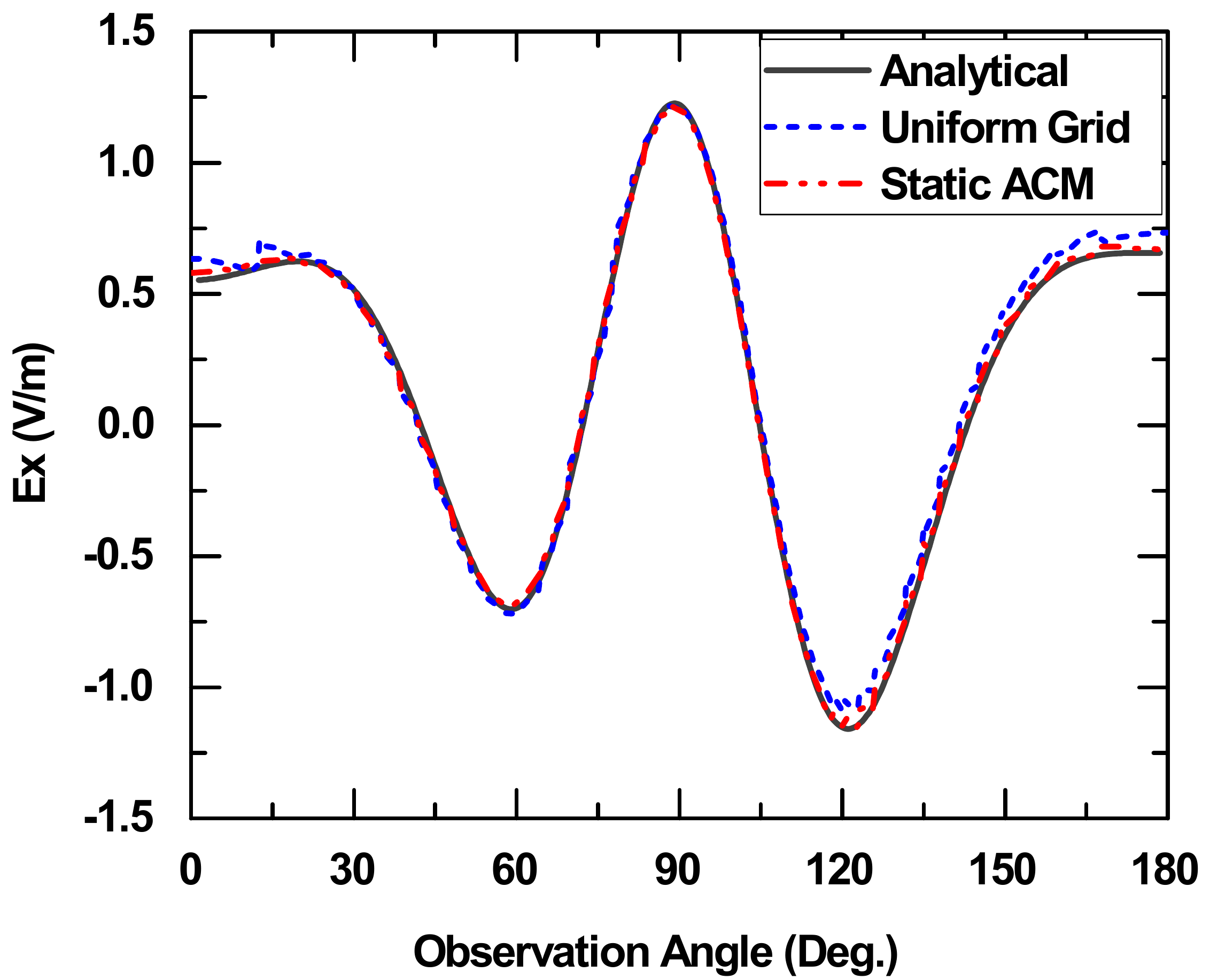}\label{fig:Sphere-yz}}
  }\caption{(a) Snapshots of the electric field $E_x$ in the 3D test at $13.3$ ns, together with the corresponding static ACM grid $h_0+2$. Comparison between the results obtained from analytical solution, uniform grid $h_0+0$, and static ACM grid $h_0+2$, in the 3D test in the (b) $xz$ plane and (c) $yz$ plane.}
  \label{fig:Sphere-E}
\end{figure}

\section{Numerical Examples}\label{sec:NumericalExamples}
In this section, numerical examples are given to first validate the implementation and show the accuracy of the DGTD method with static ACM grids in resolving objects with curved boundaries. The accuracy and efficiency of the DGTD method with dynamic ACM grids are then demonstrated through a 2D and a 3D problem. All examples with ACM grids are simulated with the LTS technique.

\subsection{Validation of the Static ACM}
To validate the DGTD solver with ACM grids in 2D and 3D, the scattering of a monochromatic plane wave by a conducting cylinder, a dielectric cylinder, and a conducting sphere is simulated, respectively. In the 2D examples, the wavelength of the incident plane wave is $\lambda_{\textrm{inc}} = 0.4$ m and the radius of the scatterer is equal to $0.1$ m. For the dielectric cylinder case, the dielectric constant in the cylinder is $\eps_{\textrm{r}} = 6$.

To resolve the wave propagating in the background, a structured mesh with a uniform size $h_0 = {\lambda_{\textrm{inc}}}/{25.6}$ is employed to discretize the free space. On top of the $h_0$ mesh, the ACM technique is applied to generate different levels of refinement at the scatterer boundary to resolve its geometrical curvature, where the first level of refinement results in mesh elements with a size of $h_1 = h_0 / 2$, the second level of refinement results in mesh elements with a size of $h_2 = h_1 / 2$, and so on. The ACM mesh with element sizes ranging from $h_0$ to $h_r$ is referred to as the $h_0 + r$ mesh hereafter. For example, the $h_0+0$ mesh denotes the uniform $h_0$ mesh in the entire solution domain. Since the refined elements are simply employed to resolve the curved boundary, and the grid does not change during the time-domain simulation, it is referred here to as the static ACM grid.

Figure \ref{fig:ConductingRod} presents the electric field distributions in the 2D conducting cylinder example, where three sets of results obtained from the analytical solution, the ACM grid $h_0+2$, and the uniformly coarse grid $h_0+0$ are shown. From the grids shown in Figs. \ref{fig:CR-2} and \ref{fig:CR-3}, it is clear that the $h_0+2$ mesh provides a much better resolution of the curved boundary. To have a quantitative comparison, the electric fields are recorded on a circle with a radius of $0.12$ m. The root-mean-square (RMS) errors of the numerical results compared to the analytical solution $\E_{\textrm{anal}}$
\begin{align}
\textrm{RMS} = \sqrt{\frac{1}{N_{\textrm{obs}}} \sum_{i=1}^{N_{\textrm{obs}}} \| \E \fun{\v{r}_i} - \E_{\textrm{anal}} \! \fun{\v{r}_i} \|^2 }
\end{align}
are presented in Tab. \ref{tab:RMS}, from which it can be seen that by decreasing the mesh size at the boundary from $h_0$ to $h_3$, the RMS error of the numerical solution decreases consistently, due to the better boundary resolution provided by the refined ACM grid.

The electric field distributions in the 2D dielectric cylinder example are presented in Fig. \ref{fig:ZIDielectrocRod}, where the results from the analytical solution, the ACM grids $h_0+3$ and $h_0+1$ are shown. Due to the dielectric constant in the cylinder, a finer grid is needed to resolve the shorter wavelength in the dielectric, which is why the $h_1 = h_0 / 2 = \lambda_{\textrm{diel}}/20.9$ mesh is used in Fig. \ref{fig:ZI-DR-3}. Obvious differences can be observed when comparing Figs. \ref{fig:ZI-DR-3} and \ref{fig:ZI-DR-1}, especially inside and around the dielectric cylinder, due to the poor representation of the cylinder boundary. When the $h_0+3$ mesh is used, a much more accurate numerical solution can be observed in Fig. \ref{fig:ZI-DR-2}. The accuracy improvement by refining the boundary grid can be seen more clearly in Tab. \ref{tab:RMS}, which validates the accuracy and effectiveness of the static ACM grid. To make a direct comparison, the EM scattering from a 3D conducting sphere with the same radius as the conducting cylinder is simulated, and the corresponding RMS errors are calculated and shown in Tab. \ref{tab:RMS}, from which a converging error is observed by refining the mesh on the curved spherical boundary.

The EM scattering from a larger 3D conducting sphere with a radius of $0.2$ m is considered. The electric field as well as the corresponding static ACM grid $h_0+2$ are shown in Fig. \ref{fig:Sphere-2D}. Clearly, the static ACM grid is able to resolve the electric field distribution well, especially near the curved spherical boundary. The electric fields obtained from the analytical expression, the uniformly coarse grid $h_0+0$, and the static ACM grid $h_0+2$ are recorded along two half circles with the radius of $0.4$ m in the $xz$ and $yz$ planes at $13.3$ ns, and the corresponding results are compared in Figs. \ref{fig:Sphere-xz} and \ref{fig:Sphere-yz}. From these two figures, it is evident that without the local element refinement around the curved boundary, the numerical results show obvious discrepancies from the analytical solution, even though the observation points are located half of a wavelength away from the curved boundary. With the employment of the ACM grid, the numerical result has a much better accuracy and is almost identical to the analytical solution.

\begin{table}[!t]\small
\renewcommand{\arraystretch}{1.3}
\caption{Comparison of RMS Errors between ACM Grids with Different Refinement Levels}
\label{tab:RMS} \centering
\begin{tabular}{c|c|c|c|c}
\hline
  ACM Mesh             &  $h_0 + 0$  &  $h_0 + 1$  &  $h_0 + 2$  &  $h_0 + 3$  \\
\hline
  PEC Cylinder         &  $0.0519$  &  $0.0394$  &  $0.0258$  &  $0.0198$  \\
\hline
 Diel. Cylinder       &  $0.1090$   &   $0.1063$   &   $0.0302$   &  $0.0271$  \\
\hline
 PEC Sphere           &   $0.0735$  &  $0.0488$   &  $0.0347$   &  $0.0277$   \\
 \hline
\end{tabular}
\end{table}

\begin{table}[!t]\small
\renewcommand{\arraystretch}{1.3}
\caption{Comparison of Computational Data Using Uniformly Dense and Dynamic ACM Grids}
\label{tab:Cylinder:Comput} \centering
\begin{tabular}{c|c|c|c}
\hline
  &  $\dt$  &  Total Num.      &  Tot. CPU          \\
  &    (ps)  &  of Elements     &  Time (sec.)     \\
\hline
 Uniform  & $4.5$ & $57496$ & $2380.0$    \\
\hline
 ACM       & $4.5 \sim 18.0$ & $14812 \sim 24910$  & $333.1$  \\
\hline
\end{tabular}
\end{table}

\subsection{Validation of the Dynamic ACM}
\begin{figure}[!t]
  \centering{
  \subfloat[]{\includegraphics[width=2.2in]{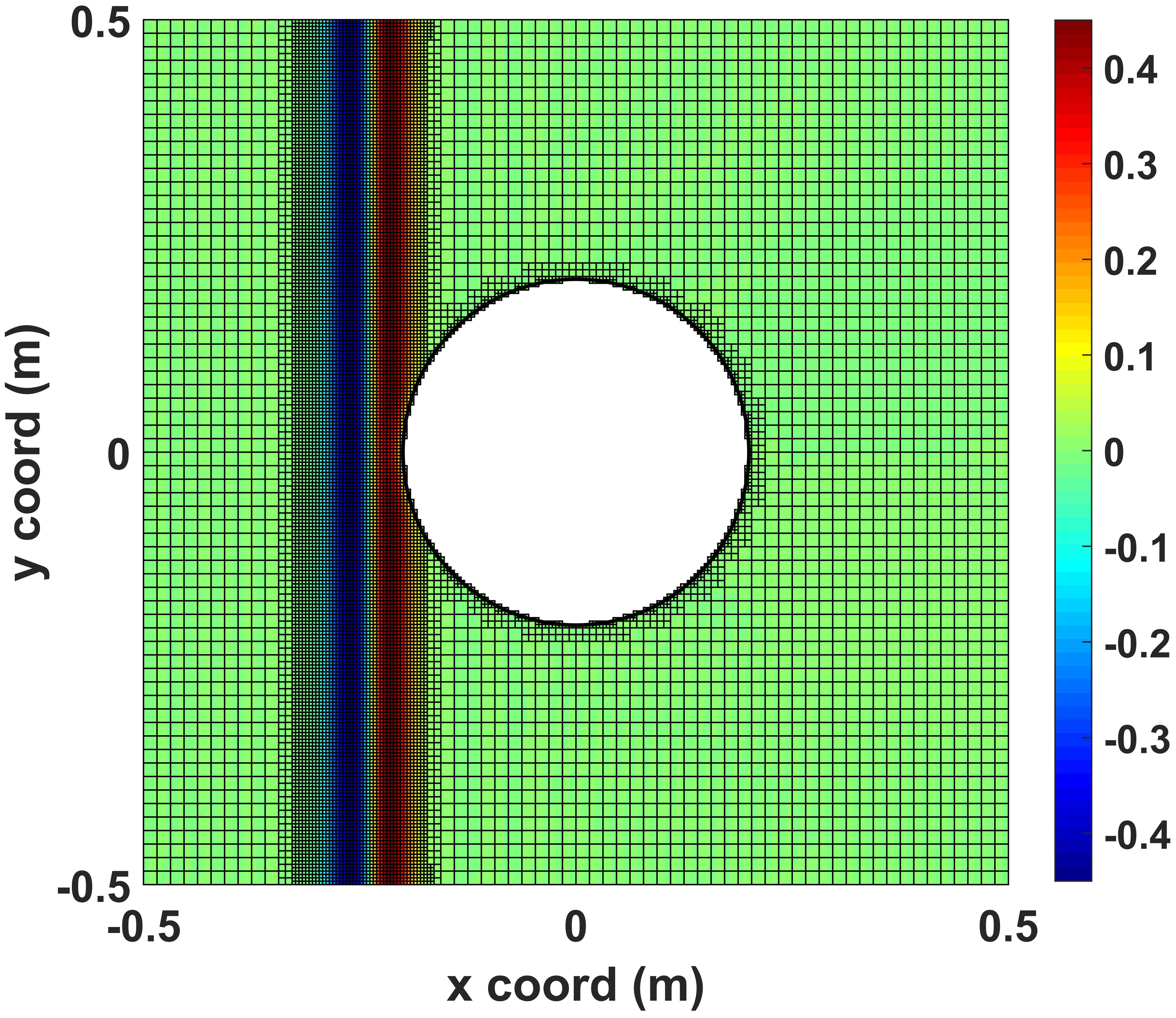}
  \label{fig:DP-1}} \hfil
  \subfloat[]{\includegraphics[width=2.2in]{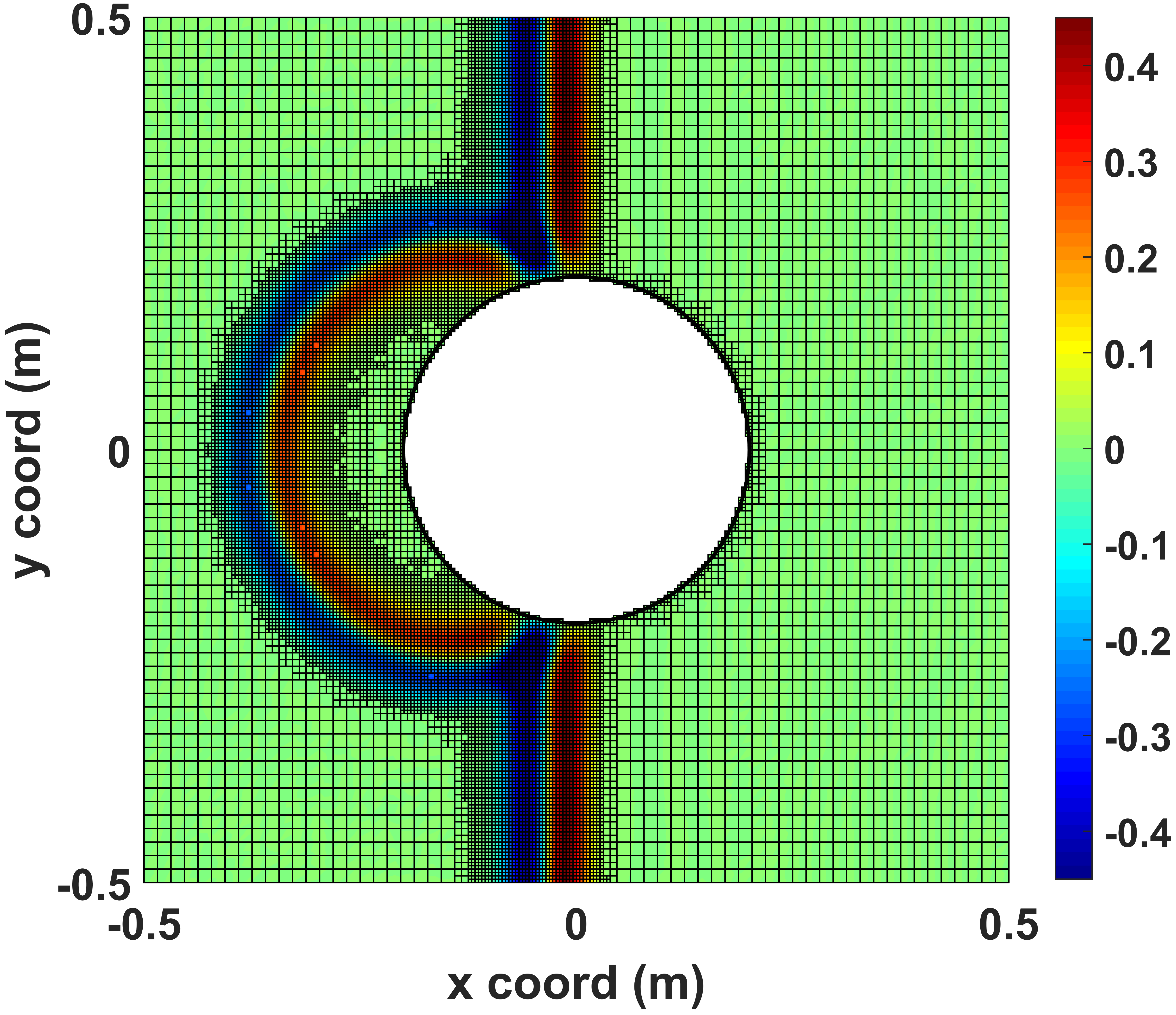}
  \label{fig:DP-2}}
  }\caption{Snapshots of the electric field $E_z$ and the ACM grid $h_0+2$ at (a) $8.55$ ns and (b) $9.23$ ns.}
  \label{fig:DACM-Pulse}
\end{figure}

\begin{figure}[!h]
  \centering{
  \includegraphics[width=2.2in]{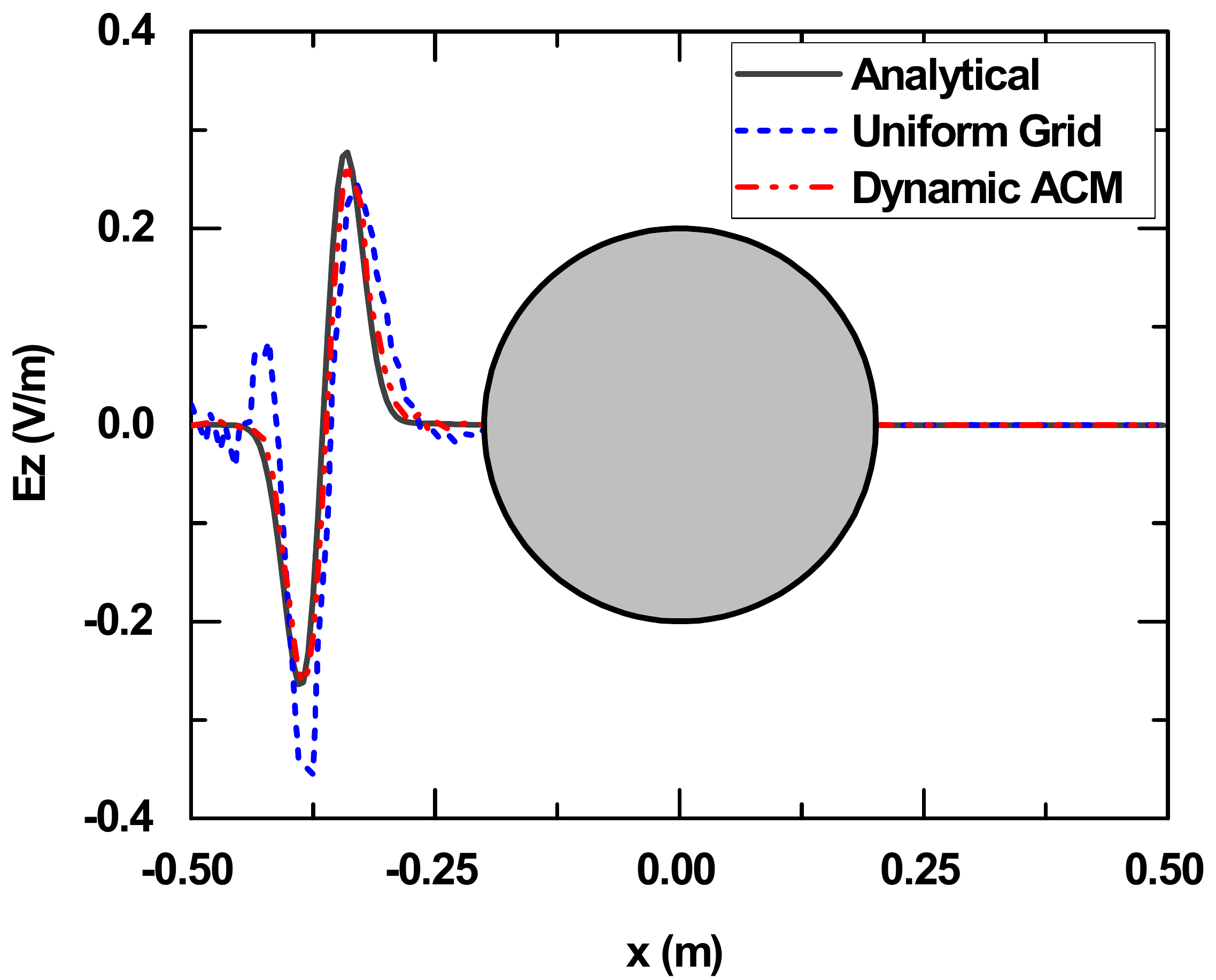}
  }\caption{Comparison between the simulated and the exact results at $9.34$ ns.}
  \label{fig:DACM-Exact}
\end{figure}

To demonstrate the accuracy and efficiency of the DGTD method with a dynamic ACM, the scattering of a plane wave with a modulated Gaussian profile from a PEC cylinder is considered. The central frequency of the incident wave is $1.50$ GHz and the pulse width is $0.53$ ns. In this example, the dynamic ACM grid $h_0+2$ is employed to capture the propagation and scattering of the highly oscillatory wave front. Shown in Fig.\!\! \ref{fig:DACM-Pulse} are the electric field distributions during the simulation at $8.55$ and $9.23$ ns, along with the corresponding dynamically refined grids. It can be seen from these two figures that the variation of the field is captured dynamically by the mesh adaptation algorithm. The electric field distribution at $9.23$ ns along the center line of the simulation domain is plotted in Fig. \ref{fig:DACM-Exact}, which shows good agreement between the analytical and the simulation results based on the dynamic ACM. Shown in Tab. \ref{tab:Cylinder:Comput} is the computational data for the simulations using a uniformly dense grid $h_2$ with a uniform time-step size and the dynamic ACM grid with the LTS technique. The computation is carried out on a computer with $18$ GB memory and the Intel(R) Xeon(R) CPU W3520 with a clock frequency of $2.67$ GHz. With the proposed method, the total computational time is reduced by more than seven times compared to the uniformly fine grid case. The DGTD with the dynamic ACM has also been applied to 3D scattering problems and the numerical experiments show a typical speedup of around 100 times as compared to the uniformly fine grid case.

It should be pointed out that although a static ACM is used to resolve the boundary curvature of the cylinder, a dynamic ACM grid can also be used very easily. In fact, the body resolution can be dynamically refined non-uniformly along the surface when and where necessary and coarsened back to the initial level when the EM pulse passes away.

\subsection{Scattering from a Cone Sphere with a Slot}
\begin{figure}[!t]
  \centering{
  \subfloat[]{\includegraphics[width=3in]{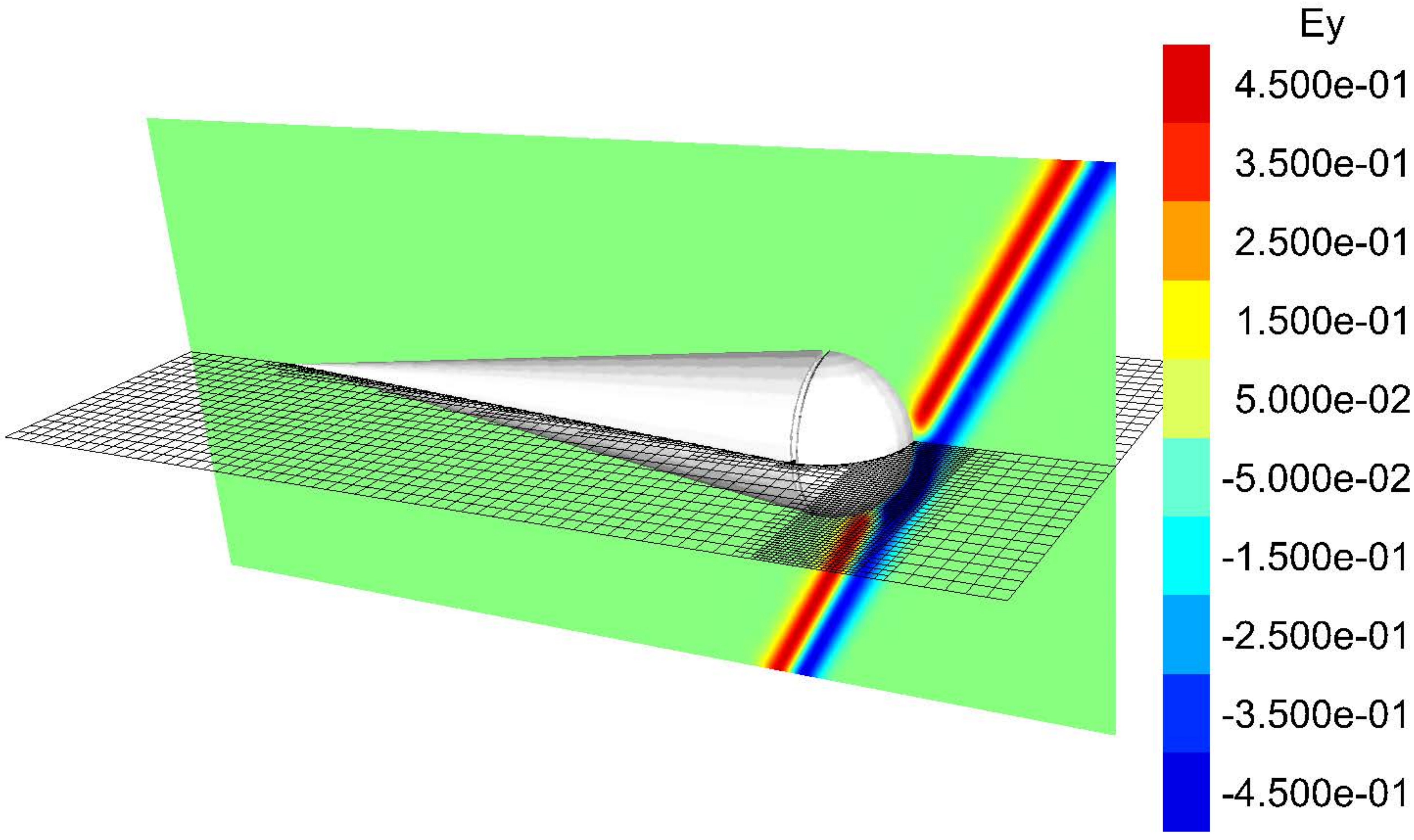}
  \label{fig:CSS-1}} \hfil
  \subfloat[]{\includegraphics[width=3in]{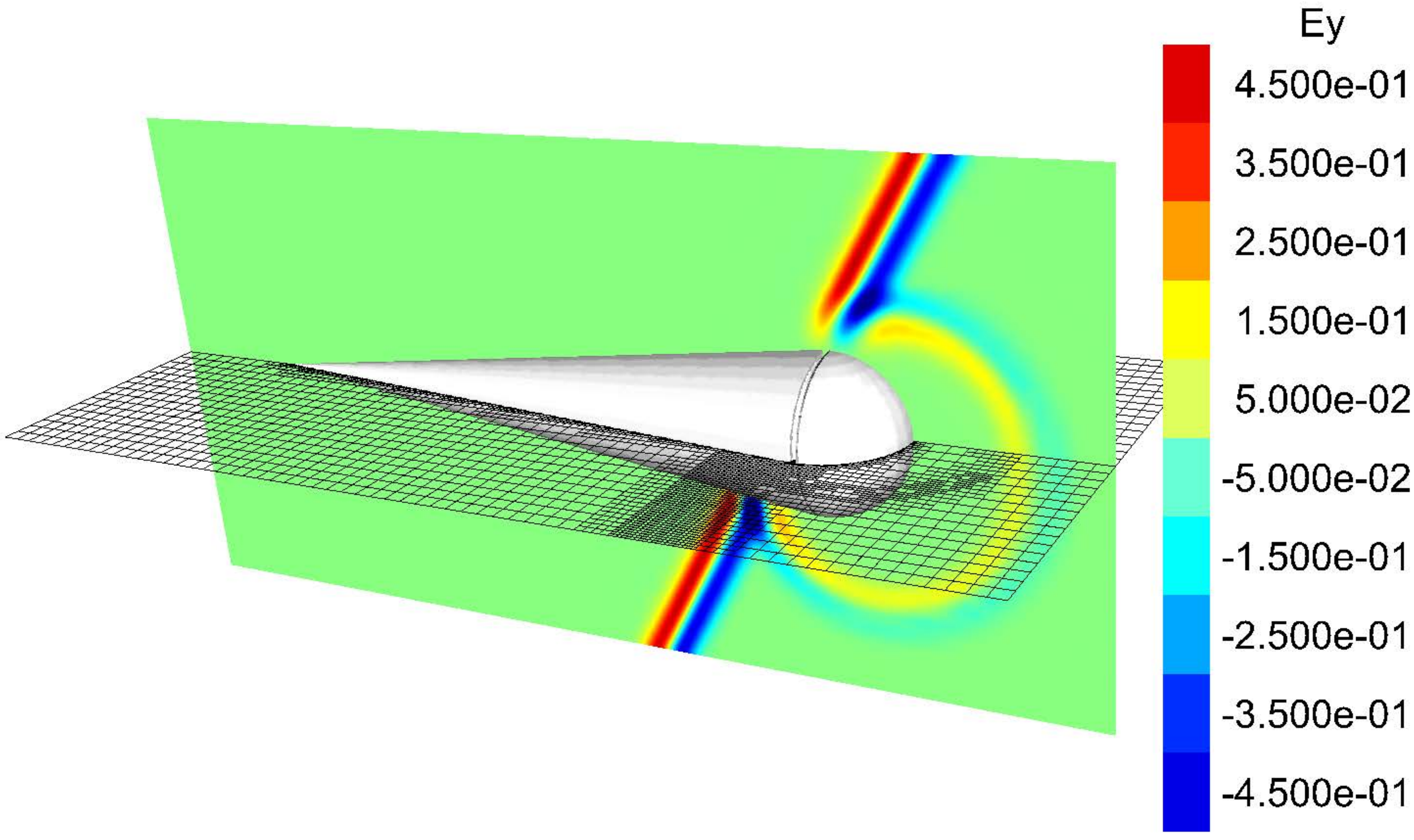}
  \label{fig:CSS-3}} \\
  \subfloat[]{\includegraphics[width=3in]{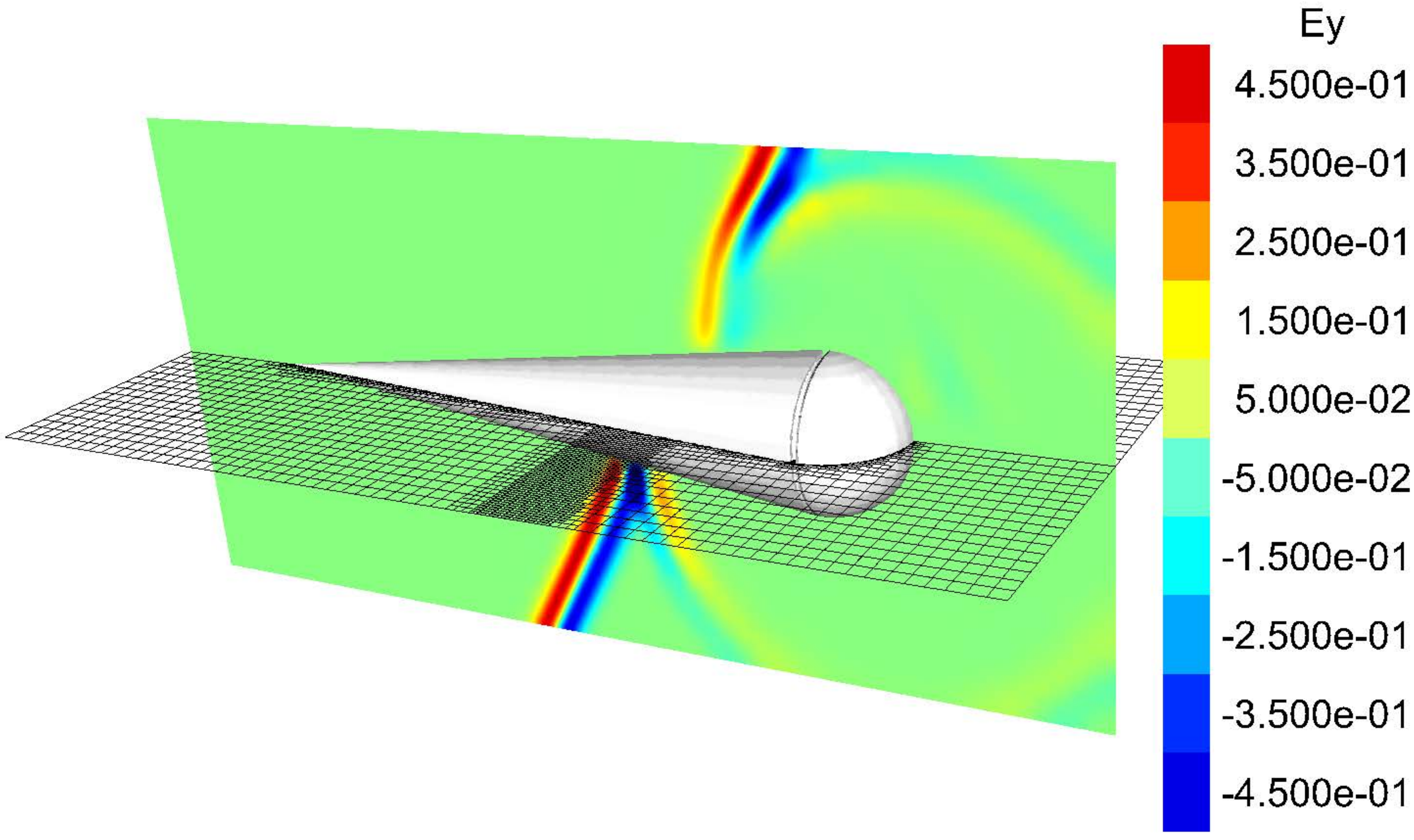}
  \label{fig:CSS-4}} \hfil
  \subfloat[]{\includegraphics[width=3in]{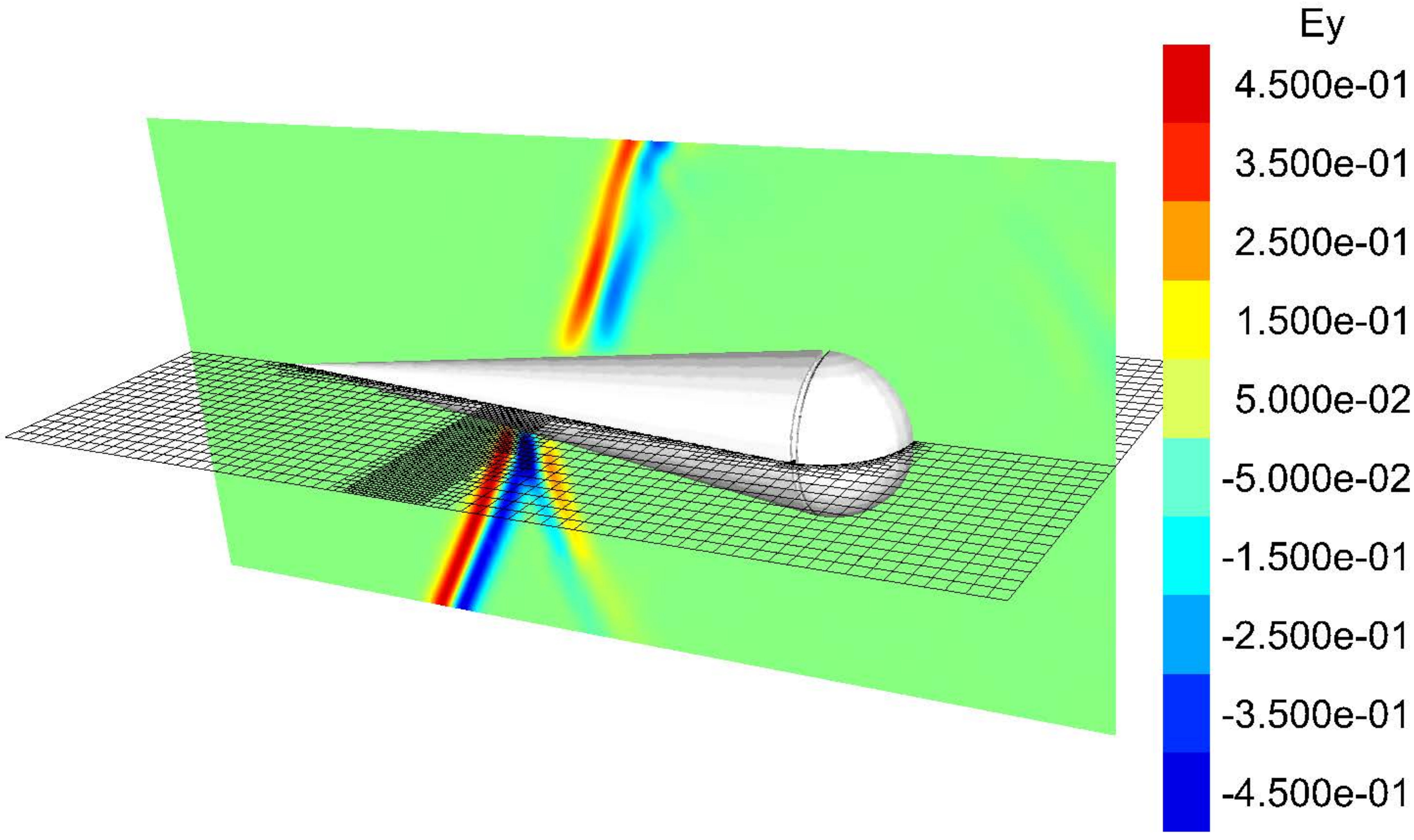}
  \label{fig:CSS-5}}
  }\caption{Snapshots of the $y$ component of the electric field and the corresponding dynamically adaptive mesh at (a) $6.75$ ns, (b) $7.50$ ns, (c) $8.50$ ns, and (d) $9.50$ ns.}
  \label{fig:ConeSphereSlot}
\end{figure}

As a more complicated 3D example, the scattering from a benchmark object, a PEC cone sphere with a slot, is considered to further demonstrate the DGTD with the dynamic ACM. Illuminated by an incident plane wave with the same Gaussian temporal profile as the one given in the preceding example, the PEC scatterer is $1.378$ m in length, and has a $1.27$-cm-wide and $1.27$-cm-deep slot around the bottom of the cone. To resolve its sharp tip and narrow slot, extremely tiny elements are required. If a uniform time-step size were applied, the total computational cost would increase dramatically. In this simulation, a dynamic ACM grid is employed by using elements with three different sizes from $h_0=\lambda_{\textrm{min}}/3.66$ to $h_2=h_0/4$. On top of the dynamic grid, the LTS technique is used to permit different time-step sizes for different elements. As a result, the simulation can be performed very efficiently, with the total number of mesh elements changing dynamically from $482482$ to $577633$ during the entire simulation, and the local time-step sizes ranging from $3.75$ to $15.00$ ps. Shown in Fig. \ref{fig:ConeSphereSlot} are the electric field distributions at $6.75$, $7.50$, $8.50$, and $9.50$ ns, together with the corresponding dynamically refined grids. Apparently, both the incident and the scattered wave fronts, where the fastest oscillations occur, can be tracked in real time, which demonstrates the capability of the proposed method.

\section{Modeling of Dispersive Media}\label{sec:Dispersive}
The electric properties of a medium are defined by its electrical conductivity and permittivity though constitutive relations, which in the simplest case have the form $\J_{\textrm{c}}=\sigma \E$ and $\D = \eps \E$. In the general case, the currents consist of the conduction current, $\J_{\textrm{c}} = -e n_{\textrm{e}} \v{u}_{\textrm{e}}$, a polarization current, and a displacement current, where $e$ stands for the charge carried by a single electron, $n_{\textrm{e}}$ stands for the electron density, and $\v{u}_{\textrm{e}}$ stands for the mean electron velocity. The first two currents are determined by the charge motion in the media, which can be described by either fluid or kinetic models.

Consider the case of EM field frequencies comparable with the collision frequency of electrons. For a cold media (electron temperature $T_{\textrm{e}} = 0$) and for small perturbations of the mean velocity, the local electron momentum transfer equation is given by
\begin{align}
\frac{\partial\v{u}_{\textrm{e}}}{\partial t} = - \frac{e\E}{m_{\textrm{e}}} - \nu_{\textrm{c}}\v{u}_{\textrm{e}}
\end{align}
where $m_{\textrm{e}}$ is the electron mass at rest and $\nu_{\textrm{c}}$ is the collision frequency. Assuming that the electron density does not vary significantly within the wave cycle, one can rewrite this equation in terms of the current density as
\begin{align}
\frac{1}{\nu_{\textrm{c}}} \frac{\partial \J_{\textrm{c}}}{\partial t} = \sigma \E - \J_{\textrm{c}}
\label{eqn:5-1}
\end{align}
where
\begin{align}
\sigma = \frac{e^2 n_{\textrm{e}}}{\nu_{\textrm{c}} m_{\textrm{e}}}
\end{align}
is the electrical conductivity of cold plasma. In the limit of slow time-varying electric field or high collisionality, $\nu_{\textrm{c}} / \omega \rightarrow \infty$, we can drop the time derivative and obtain the usual (local in time) expression, $\J_{\textrm{c}}=\sigma\E$.

\begin{figure}[!t]
  \centering{
  \subfloat[]{\includegraphics[width=2.1in]{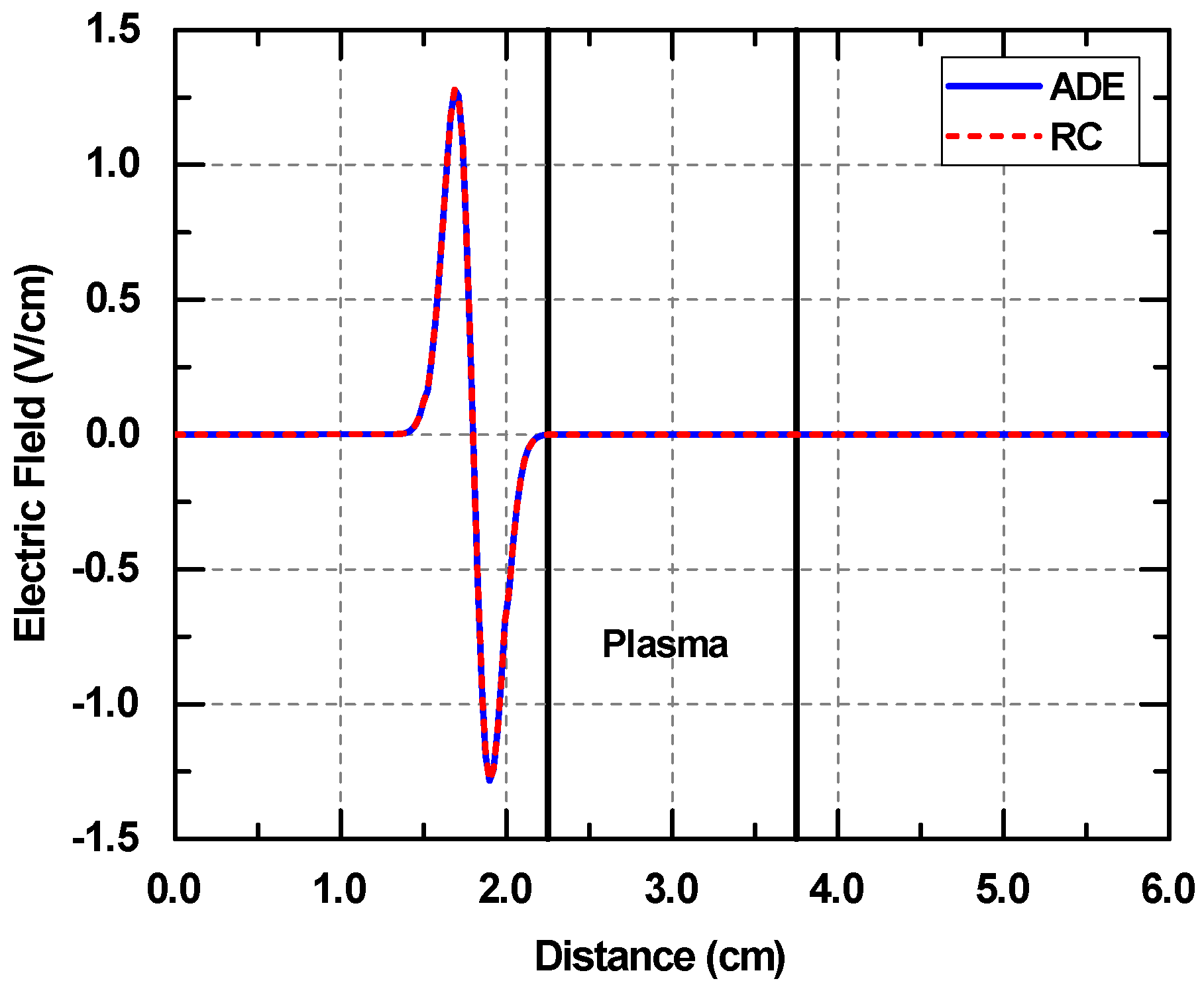}
  \label{fig:CP-1}} \hfil
  \subfloat[]{\includegraphics[width=2.1in]{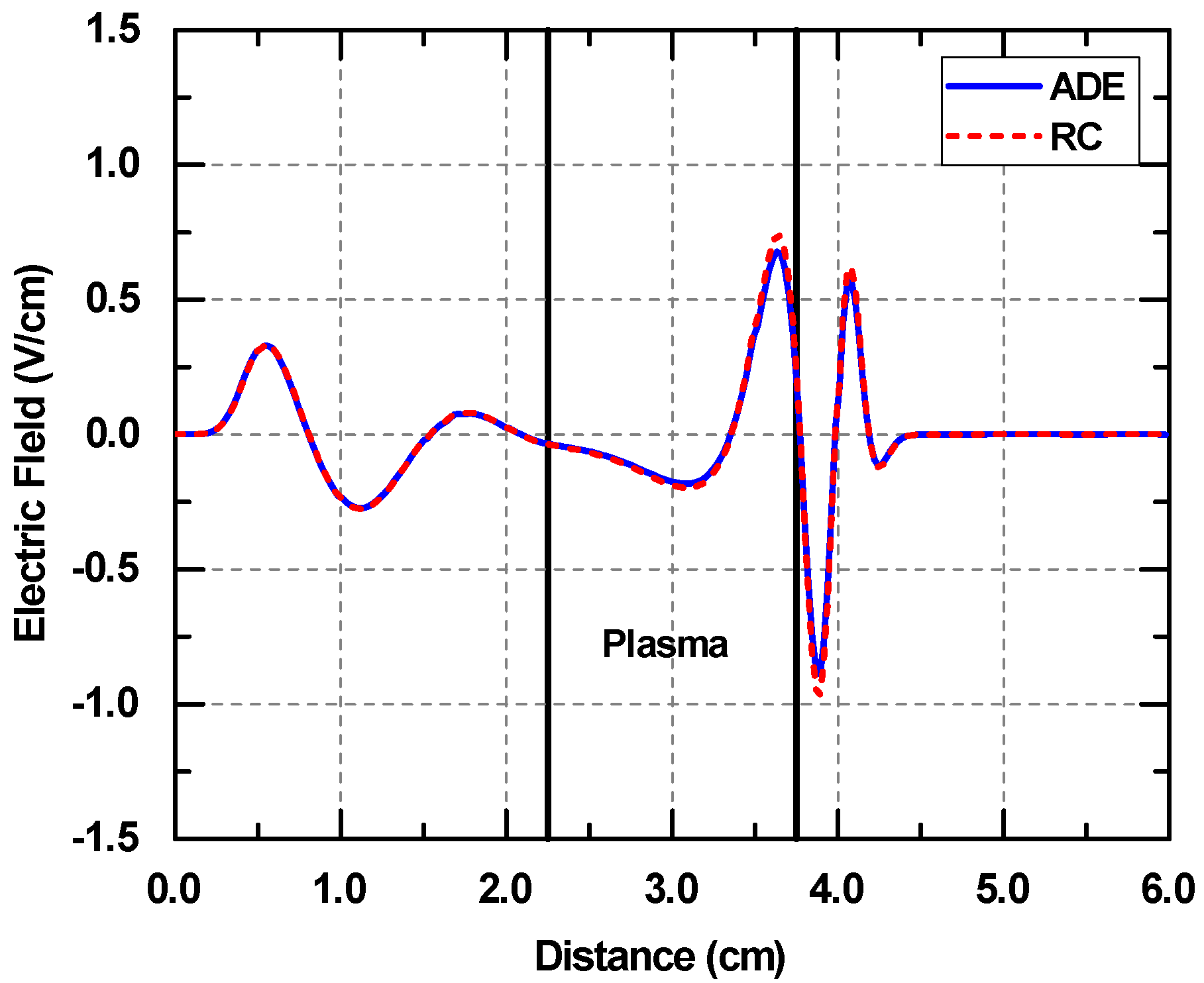}
  \label{fig:CP-2}} \hfil
  \subfloat[]{\includegraphics[width=2.1in]{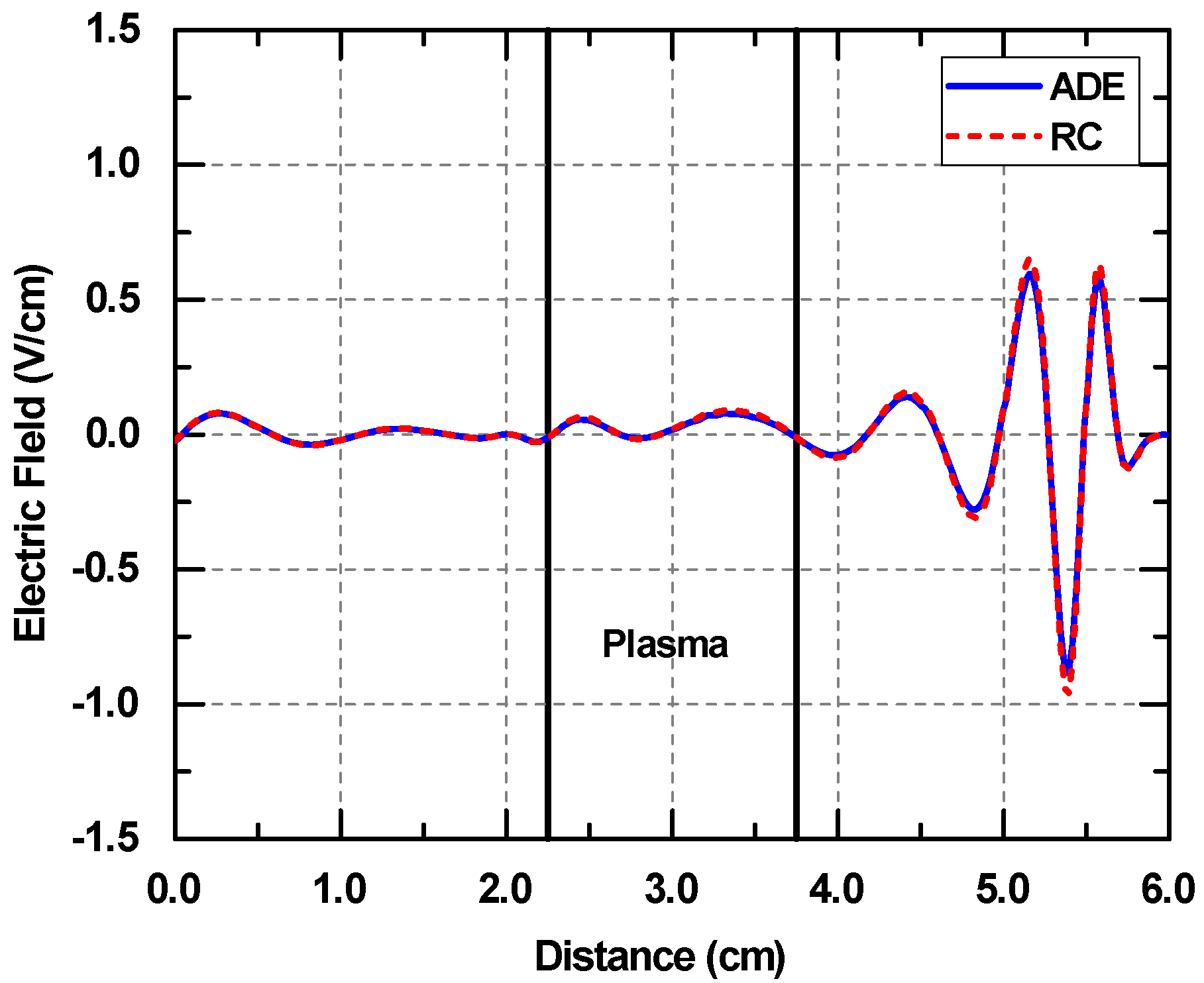}
  \label{fig:CP-3}}
  }\caption{Propagation of an initial TEM wave pulse through a cold plasma slab. Total electric field $E_y$ versus distance after (a) $1$, (b) $600$, and (c) $1000$ time steps. Comparisons are made between the numerical results obtained from the ADE and RC methods.}
  \label{fig:ColdPlasma}
\end{figure}

Equation (\ref{eqn:5-1}) for $\J_{\textrm{c}}$ can be solved in the general case by the method of time differencing
\begin{align}
\frac{1}{\nu_{\textrm{c}} \dt} \fun{ \J_{\textrm{c}}^{n+1} - \J_{\textrm{c}}^{n} } = \sigma \E^{n+1} - \J_{\textrm{c}}^{n+1}
\end{align}
which yields
\begin{align}
\J_{\textrm{c}}^{n+1} = \frac{\nu_{\textrm{c}}\dt}{\nu_{\textrm{c}}\dt + 1} \sigma \E^{n+1} - \frac{1}{\nu_{\textrm{c}} \dt} \J_{\textrm{c}}^{n}.
\end{align}
This auxiliary differential equation (ADE) method \cite{bib:Kashiwa:ADE-FDTD} for the current density has been implemented in the Runge-Kutta framework used in the DGTD-ACM code. It allows an implicit EM-charge transport coupling in the limit when spatial dispersion can be neglected. Such an implicit EM-charge transport coupling has been used in a number of works for the simulation of plasma formation during microwave gas breakdown \cite{bib:Boeuf:ADI-FDTD, bib:YAN:Plasma-Shielding}.


A recursive convolution (RC) method for dispersive media \cite{bib:Luebbers:FDFDTD} has also been implemented in the DGTD-ACM code using the high-order Runge-Kutta time integration scheme. In the RC method, the $\D$ vector is given by
\begin{align}
\D (t) = \eps_{\infty} \eps_0 \E (t) + \eps_0 \int_{0}^{t} \E (t-t') \chi (t') \textrm{d} t'
\label{eqn:RC}
\end{align}
where $\chi(t)$ denotes a susceptibility in the time domain. We have validated and compared the ADE and RC methods for a Drude model. According to the Drude model, the complex permittivity $\eps(\omega)$ for an isotropic media in the frequency domain is given by
\begin{align}
\eps(\omega) &= \eps_0 \mat{ 1 + \frac{\omega_{\textrm{p}}^2}{\omega(\textrm{j}\nu_{\textrm{c}}-\omega)} } = \eps_0 \mat{ \eps_{\infty} + \chi(\omega) }
\label{eqn:Drude}
\end{align}
where $\textrm{j}=\sqrt{-1}$ and $\chi(\omega)$ denotes a susceptibility in the frequency domain and $\omega_{\textrm{p}}$ is the plasma frequency. The Fourier transform of $\chi(\omega)$ yields a non-casual $\chi(t)$
\begin{align}
\chi(t) = \frac{\omega_{\textrm{p}}^2}{\nu_{\textrm{c}}} \mat{ 1 - \exp{\fun{\nu_{\textrm{c}} t}} }
\end{align}
which can be used in the RC method based on (\ref{eqn:RC}).

\begin{figure}[t]
  \centering{
  \subfloat[]{\includegraphics[width=2.1in]{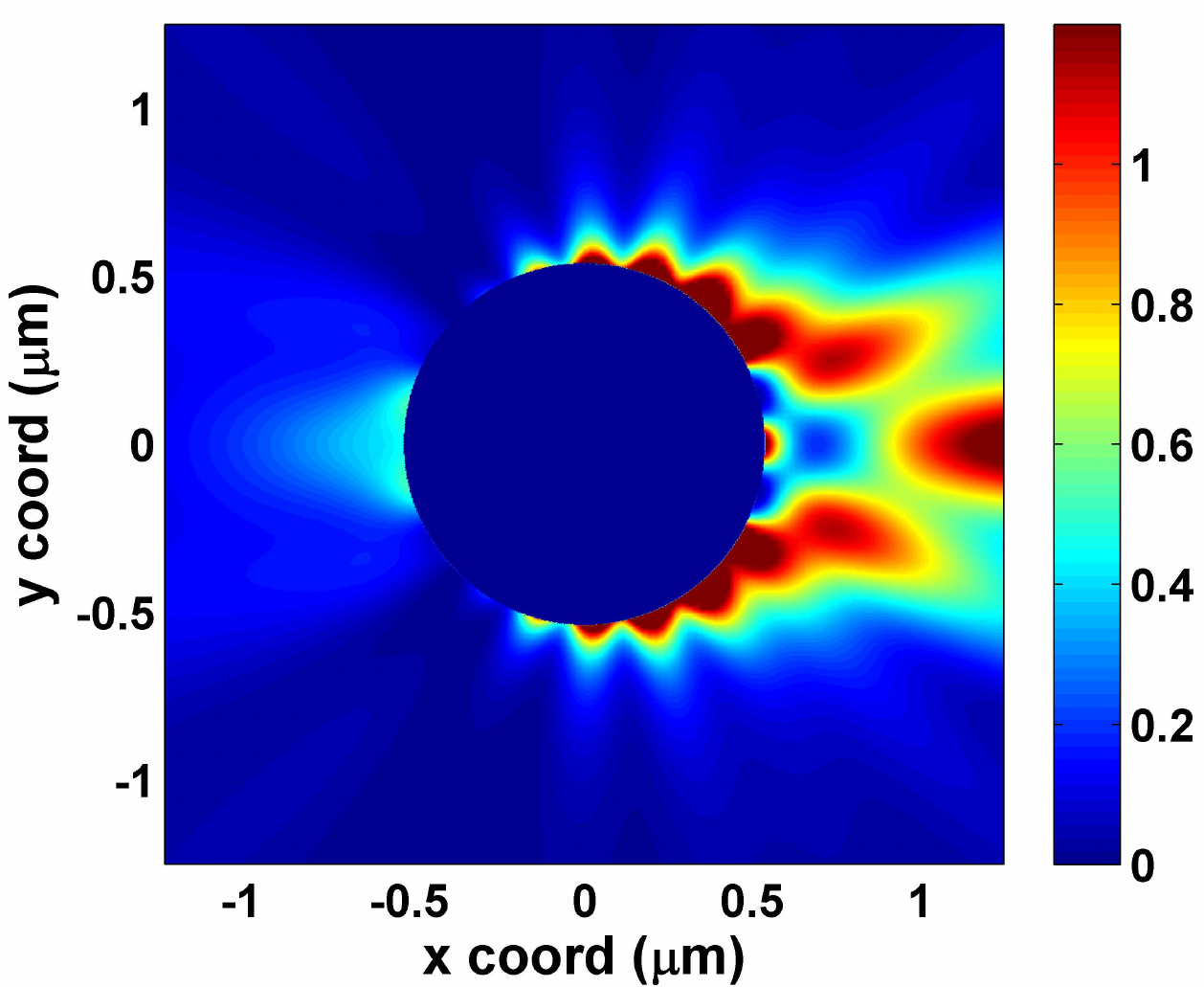}
  \label{fig:PR-1}} \hfil
  \subfloat[]{\includegraphics[width=2.1in]{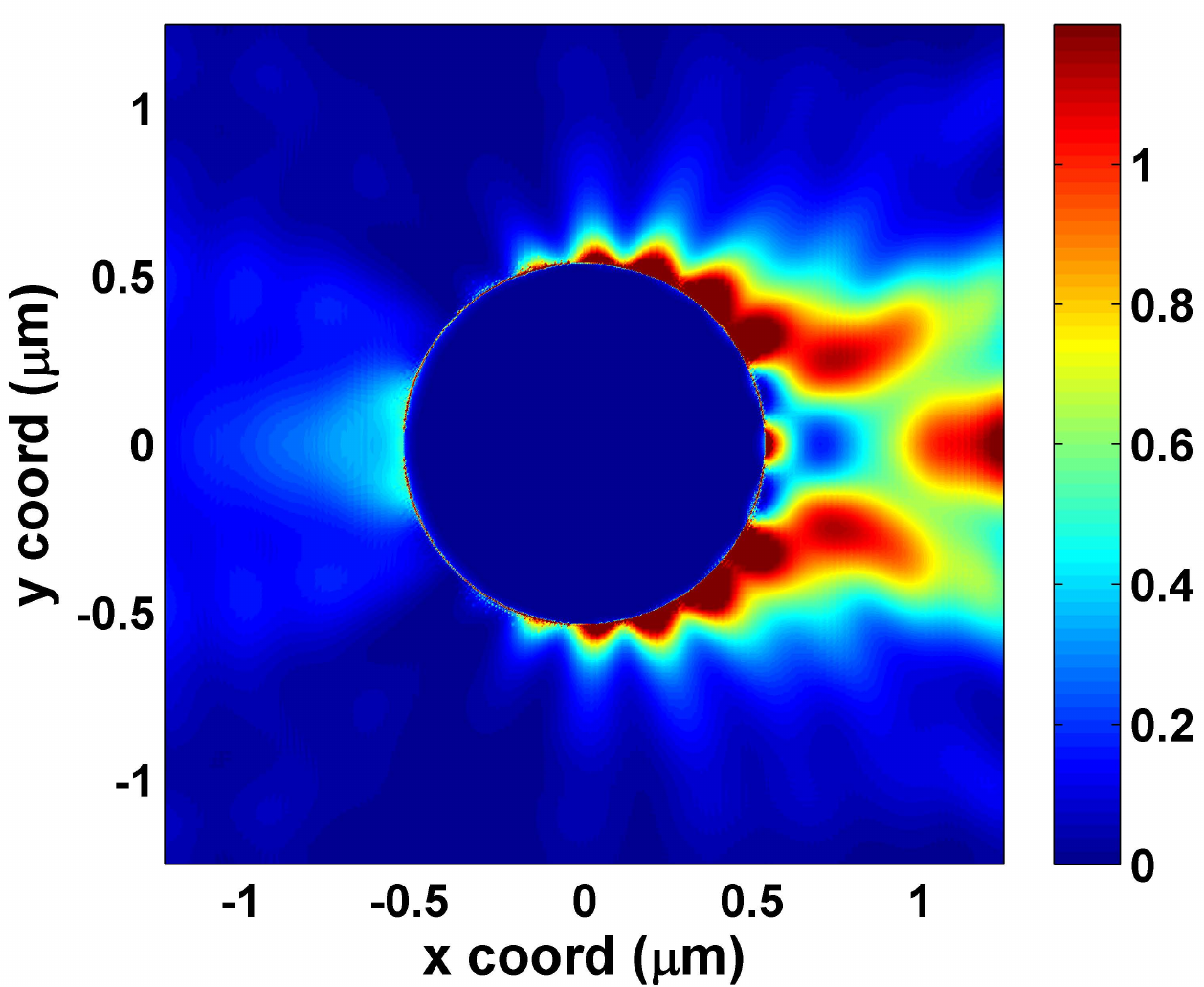}
  \label{fig:PR-2}} \hfil
  \subfloat[]{\includegraphics[width=2.1in]{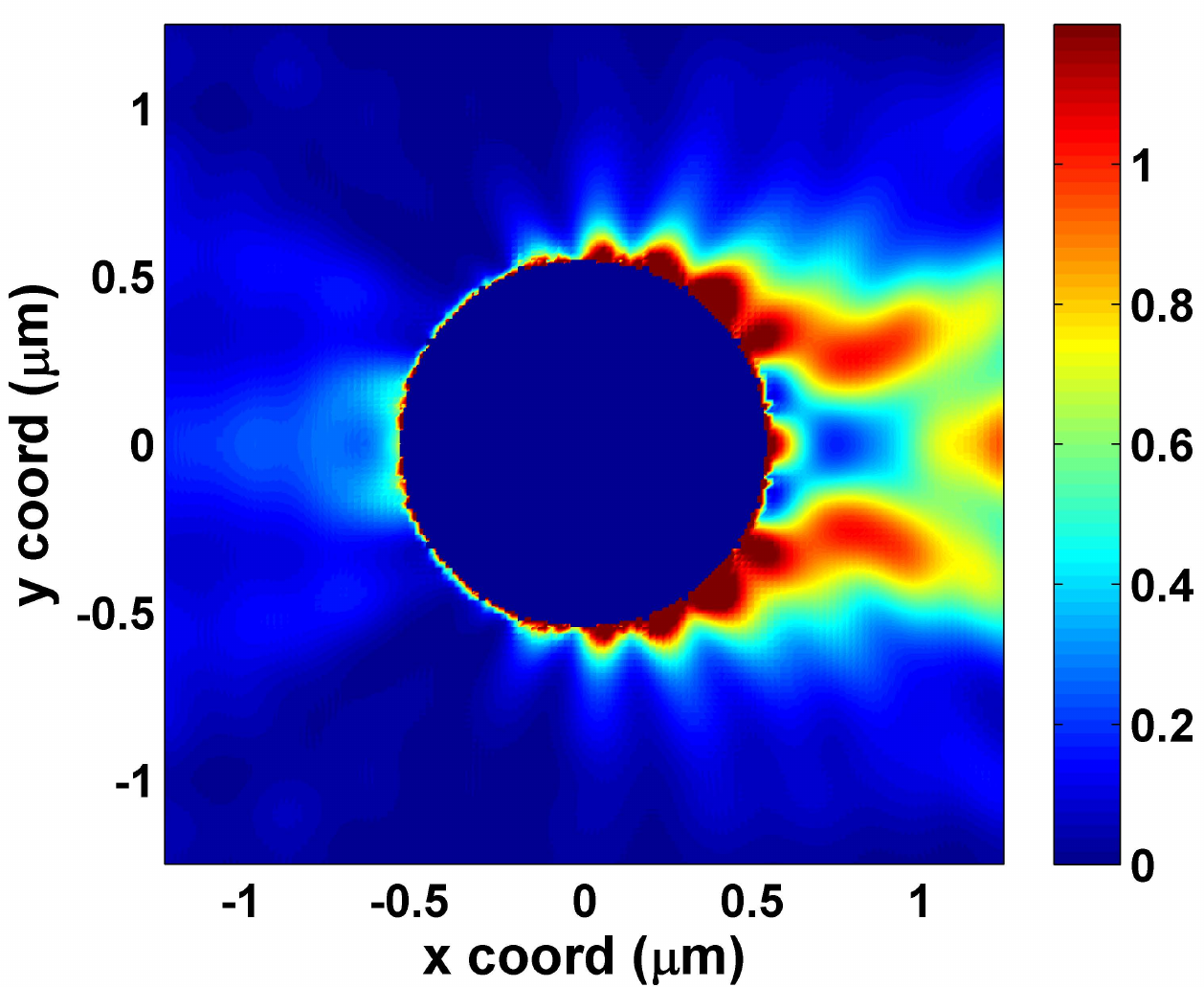}
  \label{fig:PR-3}}
  }\caption{RMS value of the electric field distributions over $30$ EM cycles. (a) Analytical result; (b) Numerical result obtained using the ACM grid; and (c) Numerical result obtained using the uniformly coarse grid.}
  \label{fig:PlasmonicRod}
\end{figure}

\begin{figure}[!t]
  \centering{
  \subfloat[]{\includegraphics[width=2.1in]{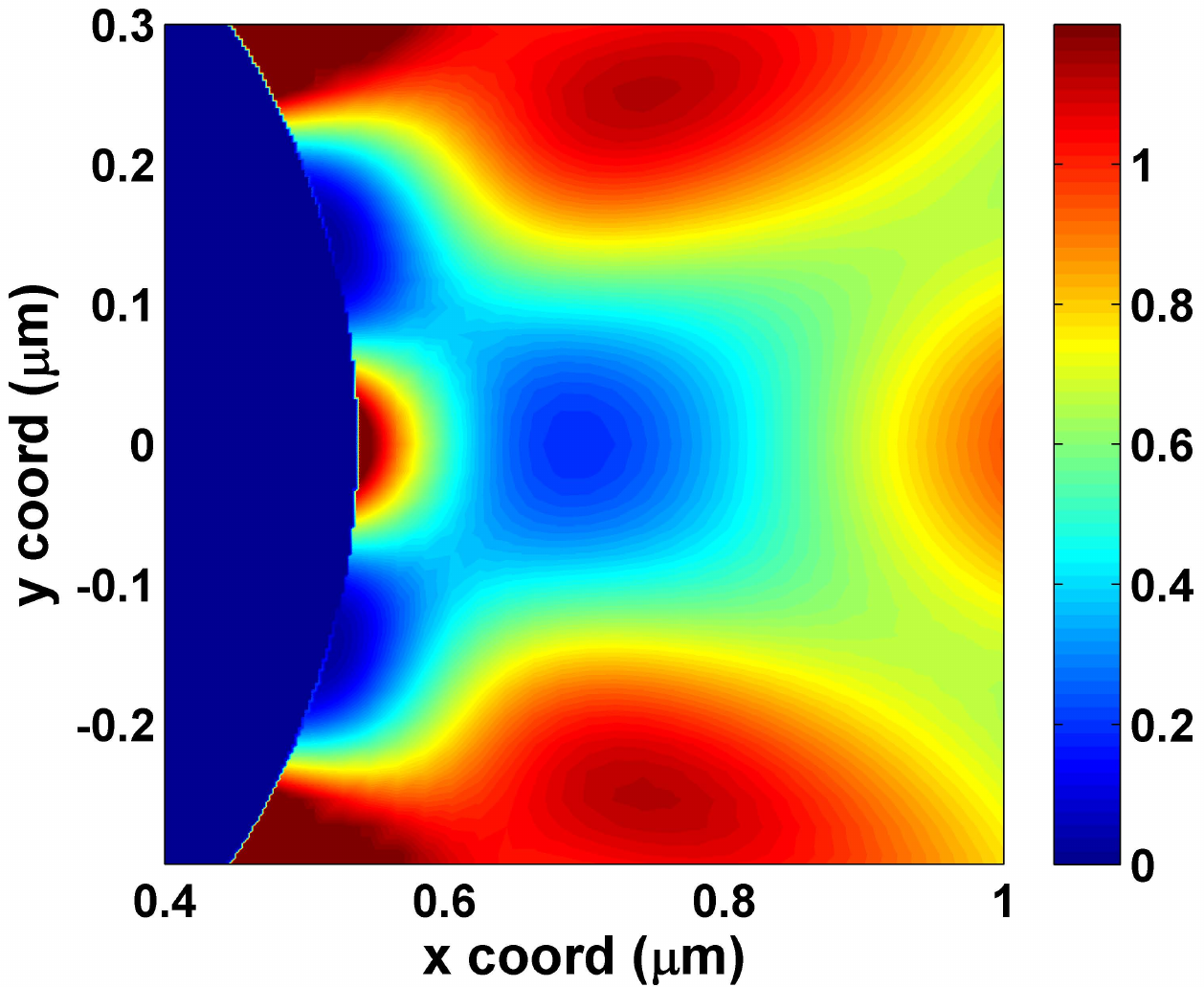}
  \label{fig:ZI-PR-1}} \hfil
  \subfloat[]{\includegraphics[width=2.1in]{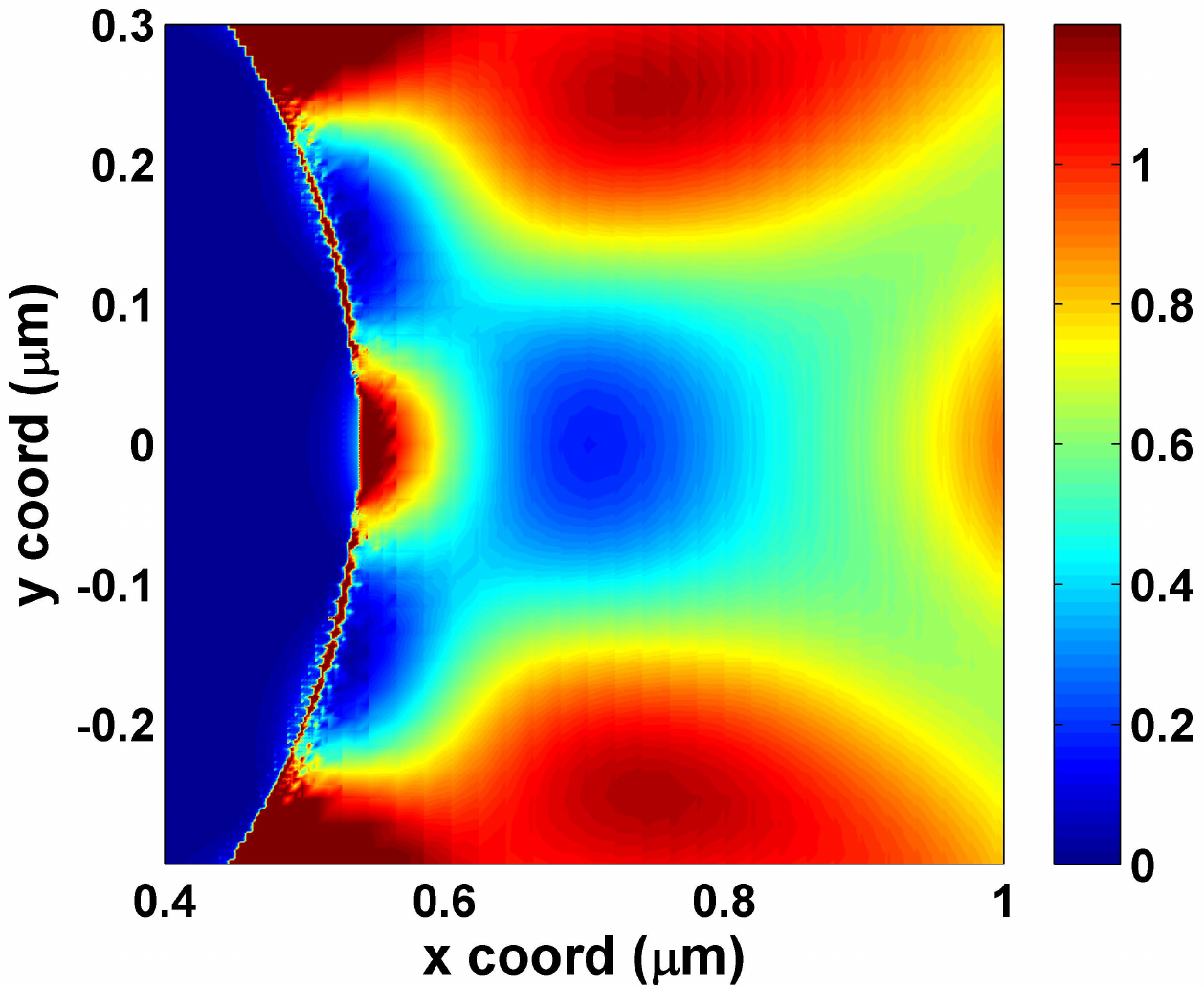}
  \label{fig:ZI-PR-2}} \hfil
  \subfloat[]{\includegraphics[width=2.1in]{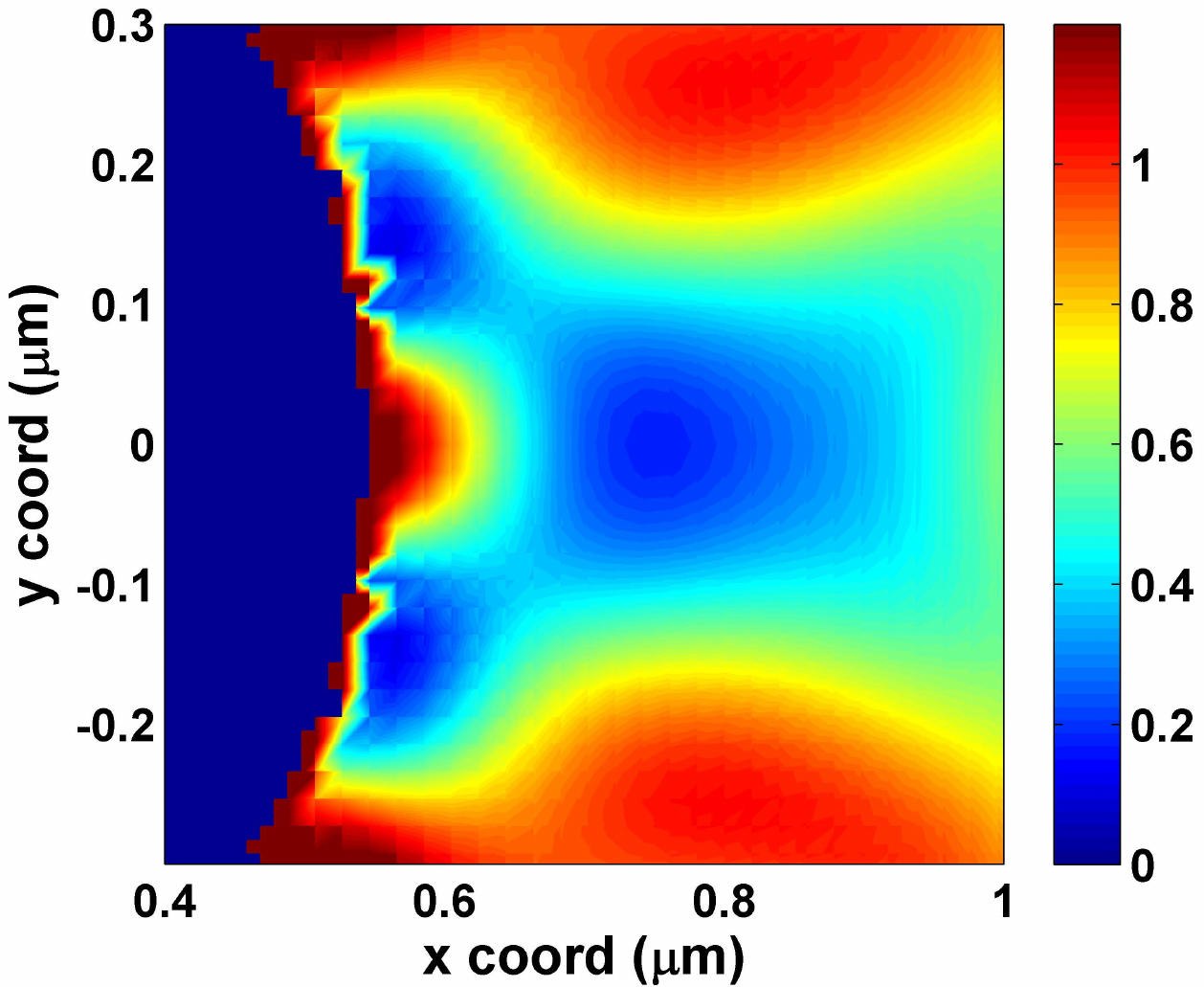}
  \label{fig:ZI-PR-3}}
  }\caption{Zoomed-in view of the region $x \in [0.4, 1.0]$ $\mu$m and $y \in [-0.3, 0.3]$ $\mu$m. (a) Analytical result; (b) Numerical result obtained using the ACM grid; and (c) Numerical result obtained using the uniformly coarse grid.}
  \label{fig:PlasmonicRod-ZoomIn}
\end{figure}

\subsection{Cold Plasma Slab}
To validate the ADE and RC methods, we calculated an EM pulse incident on a $15$-mm thick cold plasma slab. The plasma slab occupies a region from $22.5$ mm to $37.5$ mm with the computational domain length being $60$ mm. The time step is $0.125$ ps and absorbing boundaries are used at the ends of the computational domain to eliminate unwanted reflections. The plasma frequency $f_{\textrm{p}}$ is $28.7$ GHz ($\omega_{\textrm{p}}=2\pi f_{\textrm{p}}$) and the collision frequency $\nu_{\textrm{c}}$ is $2000$ GHz. For these conditions, $\nu_{\textrm{c}}/\omega \approx 10$ and thus a collisional regime is realized. In order to eliminate zero frequency incident energy, calculations are made for a normally incident plane wave with a time behavior given by the derivative of a Gaussian pulse. The spectrum of the incident pulse rises sharply but smoothly from zero frequency, peaks at approximately $50$ GHz, and is $10$ dB down from this peak at $100$ GHz (see \cite{bib:Luebbers:FDFDTD} for details). Figure \ref{fig:ColdPlasma} shows the electric field versus position after $1$, $600$, and $1000$ time steps. The characteristic ``ringing" of the plasma is readily apparent in agreement with \cite{bib:Luebbers:FDFDTD}. One can observe an excellent agreement between the DGTD results obtained using the ADE and RC methods. More complex models taking into account spatial dispersion (or high-temperature effects) \cite{bib:Yuan:HighTempPlasma} can be easily implemented.

\subsection{Scattering from a Plasmonic Rod}
One application where the dynamic ACM becomes very useful is the simulation of the surface plasmon in the vicinity of a metal-dielectric interface, where the charge density oscillations and associated electromagnetic fields are known as surface plasmon-polariton waves. The intensity of EM fields decays exponentially away from the interface. In order to capture such an exponential decrease, a very dense mesh is usually needed. With the dynamic ACM technique and the LTS scheme introduced in the preceding sections, such a phenomenon can be captured with both a high accuracy and a good efficiency. To simulate the surface plasmon-polariton wave, the metal can be modeled with the Drude model given in (\ref{eqn:Drude}).

As an example, the EM scattering from a plasmonic rod is simulated. Illuminated by a $430.501$-nm monochromatic plane wave, the plasmonic rod has a radius of $538.126$ nm, a plasma frequency of $\omega_{\textrm{p}} = 1.16 \times 10^{16}$ rad/s, and a collision frequency of $\nu_{\textrm{c}} = 1.22 \times 10^{14}$ Hz. To demonstrate the exponential decay of the EM field intensity, the RMS value of the electric field distribution is defined as
\begin{align}
E_{\textrm{rms}} = \frac{1}{T} \int_{t_0}^{t_0+T} \| \E \|^2 \textrm{d}t
\end{align}
where $T$ stands for a certain period of time, which is set as $30$ cycles of the incident EM field in this example. Shown in Fig.\! \ref{fig:PlasmonicRod} are three sets of results obtained from the analytical expression (Fig.\! \ref{fig:PR-1}), the static ACM grid case (Fig.\! \ref{fig:PR-2}), and the uniformly coarse grid case (Fig.\! \ref{fig:PR-3}). From these figures, it is clear that with the ACM grid, the numerical results match the analytical solutions very well, while the results obtained from the uniform grid have obvious discrepancy, especially near the rod boundary where the field distribution varies rapidly. The RMS errors recorded on a $700$-nm-radius circle in the uniform and the ACM grid cases are $0.0553$ V/m and $0.0300$ V/m, respectively.

\section{Conclusion}\label{sec:Conclusion}
A nodal-based DGTD algorithm with dynamically adaptive Cartesian meshes (ACM) has been developed for computation of electromagnetic fields in dispersive media. The DGTD-ACM solver takes advantages of hierarchical Cartesian grids to locally refine the non-conformal discretization elements to better represent material interfaces and curved boundaries. More importantly, the algorithm can dynamically adjust the size of each element in the real time to simulate propagation of electromagnetic pulses within the solution domain. To alleviate the time-step limitation due to the stability condition of an explicit time integrator, a local time-stepping technique is adopted to permit different time-step sizes for elements with different sizes for a better computational efficiency. Both 2D and 3D simulations of electromagnetic wave scattering and diffraction over conducting and dielectric cylinders and spheres demonstrate that the proposed method can achieve a good numerical accuracy at a reduced computational cost compared with uniform meshes. When compared with a uniformly dense mesh, the cost reduction is up to several orders of magnitude. For simulations of dispersive media, the auxiliary differential equation (ADE) and the recursive convolution (RC) methods are implemented for a local Drude model and tested for a cold plasma slab and a plasmonic rod. In future work, we plan to implement more advanced models of charge transport taking into account spatial dispersion, electron diffusion and ionization processes. The DGTD-ACM method with LTS is expected to provide a powerful tool for computations of electromagnetic fields in complex geometries for applications to high-frequency electronic devices, plasmonic THz technologies, as well as laser-induced and microwave plasmas.

\section*{Acknowledgement}
This work was supported by a Small Business Technology Transfer (STTR) grant FA8650-15-M-1940 from the Air Force Research Laboratories.

\section*{References}
\bibliographystyle{elsarticle-num}
\bibliography{IEEEabrv,./DGTD_ACM_JCP}

\end{document}

%% file: header_jcp.tex
\usepackage{array}
\usepackage[caption=false,font=footnotesize]{subfig}
\usepackage{amsmath,amssymb,amsfonts,amsthm,mathrsfs}
\usepackage{mathbbold}
\usepackage[english]{babel}
\usepackage{bm}
\usepackage{color,xcolor}
\usepackage{caption}  
\usepackage{graphicx}
\usepackage{hyphenat}
\usepackage{manfnt}
\usepackage{multicol} 
\usepackage{setspace}
\usepackage{stfloats}
\usepackage{url}
\usepackage{verbatim}
\usepackage{multimedia}
\usepackage[linesnumbered,boxed]{algorithm2e}
\usepackage{threeparttable}
\usepackage{CJK,CJKnumb}

\definecolor{UIblue}{RGB}{0,38,99}
\definecolor{UIorange}{RGB}{234,113,37}

\newcommand{\E}{\bm{E}}
\renewcommand{\H}{\bm{H}}
\newcommand{\D}{\bm{D}}

\newcommand{\J}{\bm{J}}

\newcommand{\eps}{\varepsilon}
\newcommand{\n}{\hat{\bm{n}}}
\newcommand{\eq}{&\!\!\!\!=\!\!\!\!&}

\newcommand{\cross}{\! \times \!}

\renewcommand{\v}[1]{\bm{#1}} 
\newcommand{\uv}[1]{\hat{\bm{#1}}} 

\renewcommand{\d}[2]{\frac{\textrm{d} #1}{\textrm{d} #2}} 
\newcommand{\pd}[2]{\frac{\partial #1}{\partial #2}} 

\renewcommand{\div}[1]{\nabla \!\cdot\! #1} 

\renewcommand{\vec}[1]{\left\{ #1 \right\}}
\newcommand{\mat}[1]{\left[ #1 \right]}
\newcommand{\fun}[1]{\left( #1 \right)}

\newcommand{\dt}{\Delta t}

\let\baraccent=\= 
\renewcommand{\=}[1]{\stackrel{#1}{=}} 

\theoremstyle{definition}

\theoremstyle{remark}
